\documentclass[reprint,superscriptaddress]{revtex4-1}
\usepackage{graphicx}
\usepackage{amsmath}
\usepackage{multirow}
\usepackage{tabularx}
\usepackage[hidelinks]{hyperref}
\usepackage{fancyhdr}

\pdfoutput=1
\begin{document}

\title{Examining an  Image Reconstruction Method in Infrared Emission Tomography}

\author{Loizos Koutsantonis}
\affiliation{The Cyprus Institute, Konstantinou Kavafi 20,  2121 Nicosia, Cyprus}

\author{Aristotelis-Nikolaos Rapsomanikis}
\affiliation{Physics Department, National and Kapodistrian University of Athens, Ilissia University Campus, 15771 Athens, Greece}

\author{Efstathios Stiliaris}
\affiliation{Physics Department, National and Kapodistrian University of Athens, Ilissia University Campus, 15771 Athens, Greece}
\affiliation{The Cyprus Institute, Konstantinou Kavafi 20,  2121 Nicosia, Cyprus}

\author{Costas N. Papanicolas}
\email[Corresponding author:]{cnp@cyi.ac.cy}
\affiliation{The Cyprus Institute, Konstantinou Kavafi 20,  2121 Nicosia, Cyprus}

\date{\today}

\begin{abstract}
We present and evaluate the application of the "Reconstructed Image from Simulations Ensemble" (RISE), a novel tomographic image reconstruction method, in infrared tomography. We demonstrate that established methods of photon emission tomography, widely used with penetrating ionizing radiation,  are applicable to infrared radiation. RISE, the method of choice, employs statistical physics concepts and utilizes Monte Carlo techniques to construct the imaged object from its infrared planar projections. The validity of the InfraRed Emission Tomographic (IRET) method is demonstrated, and the efficacy of RISE is evaluated with A) simulated data and B) experimental sets of infrared projections obtained from a thermal phantom with an infrared camera.  For the simulation studies presented, the reconstructed images obtained with RISE and the well - known Algebraic Reconstruction Technique (ART) and Maximum Likelihood Expectation Maximization (MLEM) method were evaluated using well-established metrics. 
\end{abstract}


\maketitle

\section{Introduction}
\label{intro}

Thermal Imaging provides a diagnostic imaging modality that can be used to visualize the temperature distribution in semi-opaque media (such as tissue and skin layer) from sets of infrared radiation measurements~\cite{ring:2012, Bad:2006}. Utilizing non-ionizing radiation this imaging modality allows non-invasive measurements of temperature fluctuations in the human body caused by an abnormal generation of heat in the examined volume of interest. These fluctuations which are related to potential abnormalities of the cells' metabolic activity can be correlated to other symptoms and medical imaging observations to indicate various pathogenies \cite{lahiri:2012}.

Currently, the technique of thermal imaging is being explored and is considered potentially valuable for a number of  medical applications including the  monitoring of chemotherapy~\cite{keyser:2000}, the diagnosis of inflammatory arthritis~\cite{ring:1970}, osteoarthritis \cite{glehr:2011,romano:2011,varju:2004,mayr:1995} and the assessment of peripheral circulation~\cite{vardasca:2012}. Much effort has been extended for the detection of malignant diseases~\cite{herman:2011,santa:2009,maillard:1969} and in particular for breast cancer~\cite{kirubha:2015, kennedy:2009, ng:2008, lapa:1973}. Melanomas can be identified and localized within the skin layers of the patient by examining the different patterns of temperature distribution~\cite{ring:2012}. Various studies report that the heat generated due to the presence of cancerous lesions within the body can be up to ten times higher than that generated by healthy tissue~\cite{Pirtini:2010,Song:1984}. This deviation is caused by the abnormally higher metabolic activity of the malignant cells~\cite{kirubha:2015}. 

 Thermal imaging, despite the promise, has not been yet accepted as a technique which can be used to provide accurate and reliable diagnostic information. The main limiting factor derives from the fact that the infrared radiation emitted from the inner organs is diffused, absorbed and scattered by the intermediate tissues before exiting the skin surface. As a result, the obtained infrared images from a source below a certain depth in the body are blurred and of low-resolution~\cite{snyder:2000}. The results mentioned above highlighting the promise of thermal imaging as a medical diagnostic tool refer to single planar imaging (thermography). The obvious next step for further development that of going from planar imaging to tomographic imaging has hardly been explored presumably because of the same reasons.  Few successful attempts utilizing the traditional image reconstruction algorithms of emission tomography have been reported for 3D reconstruction of the flame temperature distribution~\cite{Hossain:2013, Goyal:2014}. In these studies, the tomographic images were produced in "idealized" conditions using static infrared projections acquired in the absence of an intermediate absorber or scatterer.  To date, the tomographic image reconstruction of the temperature distribution in an absorbing medium from the set of its infrared image projections remains a challenging problem with untapped potential. 

\section{Objectives}
\label{objectives}
The overarching objective of the work presented here is to demonstrate that infrared emission tomographic imaging is feasible and promising when implemented with the “Reconstructed Image from Simulations Ensemble (RISE)” technique. In particular this manuscript endeavors to:
\begin{enumerate}
\item[I]Extend the well established and widely used methodology of emission tomography with ionizing radiation (Positron Emission Tomography (PET) or Single Photon Emission Computed Tomography (SPECT)) to Infrared Radiation Emission Tomography (IRET) utilizing infrared radiation. Enable the extension of IR imaging from the single thermal – planar imaging ( thermography) to a tomographic modality visualizing the 3D temperature distribution in an absorbing and scattering medium.
\item[II]For the first time reconstruct tomographic images from hotspots embedded in a semitransparent medium. 
\item[III] Evaluate the success of implementing the proposed methodology employing the well established tomographic reconstruction techniques ART, MLEM, and RISE by utilizing software and hardware phantoms. Quantify the quality of the constructed tomographic images using well-accepted metrics and demonstrate that RISE is best suited for IRET.
\end{enumerate}

The argumentation and the results supporting the achievement of the objectives as mentioned above are presented as follows: the fundamental aspects of the IRET problem are reviewed, and the mathematical foundations of IRET as an inverse tomographic problem are presented in Section~\ref{IRET Methodology}~("{IRET Methodology}"). The  MLEM, the ART, and the  RISE tomographic reconstruction methods are introduced in Section~\ref{Image Reconstruction}~("{Image Reconstruction}"). In Section~\ref{Simulation Studies}~("{Simulation Studies}"), software phantoms are utilized to benchmark the performance of the various reconstruction methods in IRET along with the appropriate metrics quantifying the quality of the reconstructed images. In Section~\ref{Thermal Phantom}~("{Thermal Phantom}"), an elaborate hardware phantom is employed to showcase IRET  in achieving the stated objective. A summary of the presented results and an assessment of achieving the stated objectives along with suggestions for future work is presented in Section~\ref{Conclusions}~("{Conclusions}").

\section{InfraRed Emission Tomography (IRET) Methodology}
\label{IRET Methodology}
The IRET problem concerns the reconstruction of tomographic images depicting the temperature distribution of the imaged structure derived from its infrared planar projections obtained at different angles. 

IRET can be formulated as an inverse proble by assuming that the temperature distribution in the tomographic plane is  represented by the 2D function $T(x,y)$ and that the intensity of the radiative power per unit wavelength ($\lambda$)  emitted from a point $(x,y)$   can be calculated through the Planck's formula \cite{lienhard:2013, modest:2013}:
\begin{equation}
I(T(x,y),\lambda,x,y) = \epsilon(\lambda) \cdot \frac{c_1}{\pi \lambda^5} \cdot ({exp(\frac{c_2}{\lambda T(x,y)})-1})^{-1}
\label{Eq:Radiation}
\end{equation}
where   $\epsilon(\lambda)$ is the wavelength depended emissivity of the imaged material and $c_1=3.7417 \cdot 10^{-16} Wm^2$, $c_2 = 1.4387 \cdot 10^{-2} mK$ are the first and second 
radiation constants respectively.

\begin{figure}[ht!]
\begin{center}
\includegraphics[width=\columnwidth]{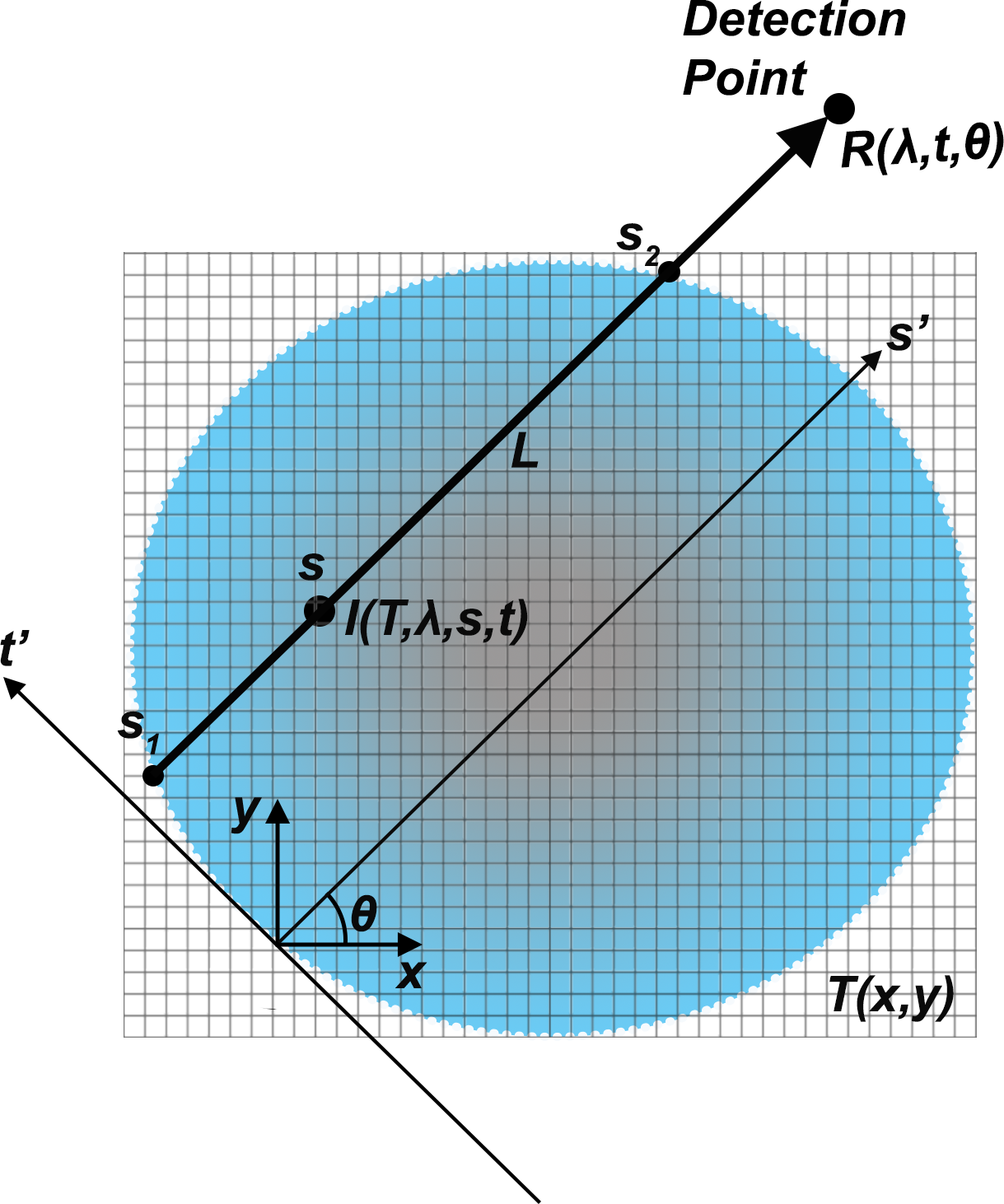}
\caption{Diagram showing the geometry of the infrared tomographic problem. The total intensity of radiative energy per unit wavelength $R(\lambda,t, \theta)$  detected by an infrared camera  is equal to the line integral of the intensity of the radiative energy per unit wavelength $I(T,\lambda,s,t)$ emitted from the points $(s,t)$ lying on the line path ($s_1,s_2)$ and attenuated along the line path ($s,s_2$). } 
\label{fig:schem}
\end{center}
\end{figure} 

\begin{figure}[ht!]
\begin{center}
\includegraphics[width=\columnwidth]{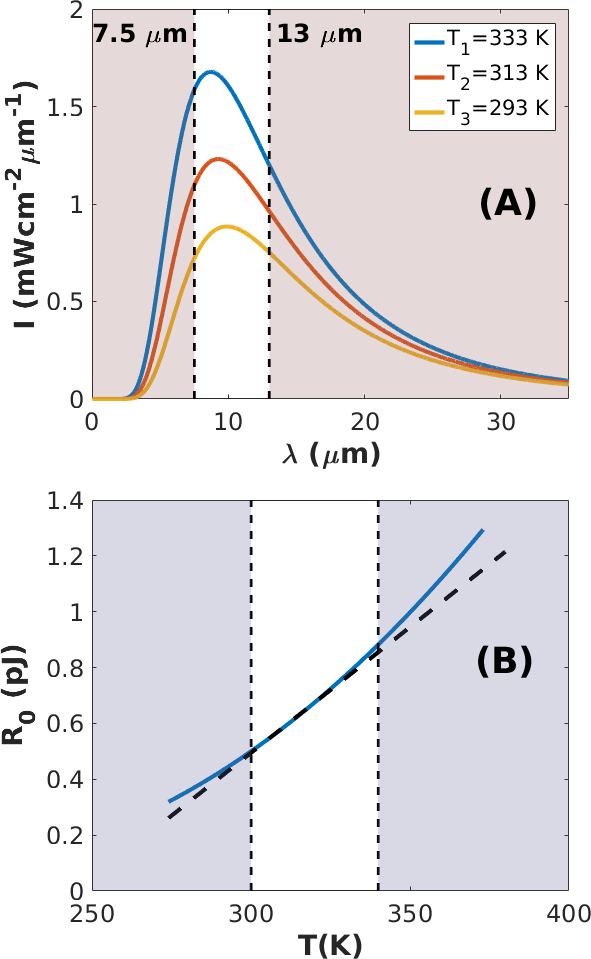}
\caption{(A) The intensity of the radiative power as a function of the wavelength plotted for different values of temperature. (B) The continuous curve indicates the total  radiative energy $R_0$ emitted in the spectral window ($\lambda_1$ = 7.5 $\mu m$,~ $\lambda_2$ = 13.0 $\mu m$) and detected by a detector of 30$\times$30 $\mu$m$^{2}$ size (typical size of thermal camera sensor) within a time window of 10 ms (typical exposure time of a thermal camera). $R_0$ is plotted as a function of the temperature $T$.   The dotted line provides the linear approximation of $R_0$ in the temperature range (300 K- 340 K).  } 
\label{fig:radiance}
\end{center}
\end{figure} 
 
Utilizing a more convenient coordinate system $(t,s)$ :
\begin{align}
t=x\cdot \cos(\theta)+y\cdot sin(\theta)  \\
s = -x\cdot \sin(\theta)+y\cdot cos(\theta)
\end{align}
where $\theta$ is the angle of the camera rotation,  the intensity of the radiative power per unit wavelength which is transmitted along the line path $L$ and being detected at $(t,\theta)$  by the detection unit is given by:
\begin{equation}
R(\lambda,t, \theta) = \int_{s_1}^{s_2} I(T(s,t),\lambda,s,t) ~e^{\bigg({-\int_s^{s_2}{k(\lambda,s',t)ds'}\bigg)}}~ds
\label{Eq:Radiation}
\end{equation}
where $k(\lambda,s,t)$ is the wavelength dependent attenuation coefficient of the imaged object.
The geometry of the above problem is best explained with the help of  Figure \ref{fig:schem}.

For an infrared detector (camera) operating in the spectral range ($\lambda_1$,~ $\lambda_2$), the total  radiative energy  being detected is:
\begin{equation}
R_0(t,\theta) = \Delta \tau  \Delta S  \int_{\lambda_1}^{\lambda_2} R(\lambda,t, \theta)~d\lambda
\label{Eq:Radiation}
\end{equation}
where $\Delta \tau$ is the  exposure time and  $\Delta S $ is the area of the camera's sensor.

For the infrared window ($\lambda_1$ = 7.5 $\mu m$,~ $\lambda_2$ = 13.0 $\mu m$) and the temperature range (300 K - 340 K) examined in this work, the relation between the  detected radiative energy $R_0$  and the  temperature distribution $T(x,y)$ of the imaged object can be approximated by a linear function (Figure \ref{fig:radiance}). We need to highlight that the linearization of the emitted radiative energy $R_0(t,\theta)$ as a function of temperature allows the formulation of IRET as a conventional emission tomographic problem.   The   IRET problem can then be formulated as a system of equations described by a projection matrix. In such a formulation of the infrared tomographic problem, a bin measurement $R_i$ corresponding to the total energy transferred by a finite width beam  can be estimated from the sum:
\begin{equation}
R_i = \sum_{j=1}^{N^2}{P_{ij}F_j}
\label{Eq:Atten_Proj}
\end{equation}
where, $F_j$ is the vector representation of the tomographic image of the temperature distribution, $N^2$ is the number of image elements,  and $P_{ij}$ is the projection matrix weighting the contribution in transmitted photons of the $j^{th}$ image element to the $i^{th}$ bin (pixelated) measurement.

In emission tomography \cite{Wernick:2004, Natterer:2001}, the Projection Matrix $P_{ij}$ not only provides the geometric weight with which the $F_j$ element (pixel/voxel) contributes to $R_i$, it also incorporates all the attenuation and scattering effects that the emitted photons undergo along their path to the detector. Unlike the x-ray or $\gamma$-ray regions, in the infrared region, the dominant light-tissue interaction is elastic scattering, and the ability to accurately incorporate its effects is important.  The incorporation of attenuation and scattering into $P_{ij}$ may entail detailed and cumbersome cascade calculations. In this generalized approach, the association of $R(\lambda,t,\theta)$ to $I(T,\lambda,s,t)$ via the Projection Matrix can no longer be interpreted in terms of simple geometrical optics as in Figure \ref{fig:schem}, but rather as the response of the detector element to the source of radiation regardless of how the photon reached it (directly or via scattering).    Methods for modeling light propagation in tissue ~\cite{ishimaru:1978, bohren:2008, Nissila:2005} are well developed, and the methodology above allows their implementation for specific cases. In real samples (e.g. in infrared tomographic medical imaging),  knowledge about the morphology (structure) and the optical properties (attenuation and scattering cross sections) will be taken into account for the construction of the Projection Matrix. 

In the examples presented in this methodological paper, we examine cases where uniform attenuation suffices. In such cases, 
\begin{equation}
{P}_{ij}= \tilde{P}_{ij}~exp(-k_u  r_{ij})
\label{Eq:Attenuation_Proj}
\end{equation}
where $\tilde{P}_{ij}$ is the "bare" Projection Matrix which takes into account only the geometrical weight of each pixel/voxel, $k_u$ is a uniform   attenuation coefficient characterizing the imaged medium, and,
$r_{ij}$ is the distance of the $j^{th}$ image element from the boundaries of the imaged volume, measured in the direction towards the $i^{th}$  pixelated detector of the thermal camera system.

    The simplified uniform attenuation map employed in this work provides a good approximation of the attenuation effects characterizing a homogeneous medium; such is the case of the hardware phantom examined in this study. Different and more complex attenuation and scattering models can be employed if knowledge on the morphology of the medium is available, as it is expected to be the case in applying IRET to medical imaging.

\section{Image Reconstruction}
\label{Image Reconstruction}

\subsection{Iterative Reconstruction Techniques}
Two widely used reconstruction techniques are employed in this study to solve the image reconstruction problem defined in Equation \ref{Eq:Atten_Proj} and provide reference images for comparison are the Algebraic Reconstruction Technique (ART) and the Maximum Likelihood Expectation Maximization (MLEM) method.
 ART \cite{Gordon:1970,Gilbert:1972} and MLEM~\cite{Shepp:1982, Lange:1984} are well established and widely used methods in emission tomography, although MLEM invariably yields superior results and it is widely regarded as representing the "State of the Art" method.  
 Both methods are capable of incorporating the projection model of Equation \ref{Eq:Atten_Proj} and can provide reconstruction results using sets containing a limited number of planar projections~\cite{Wiez:2010}. In the present work,  the Newton-Raphson variant of ART~\cite{Angeli:2009} and an accelerated version of MLEM~\cite{Hwang:2006} were used for the reconstruction of the tomographic images.

\subsection{The Reconstructed Image from Simulations Ensemble (RISE)} 
 
 RISE provides an alternative method to ART and MLEM for tomographic image reconstruction. Based on the AMIAS framework \cite{Alexandrou:2015,Markou:2018,Papanicolas:2012}, RISE produces images by parametrizing the physical characteristics of the imaged "target" and simulating a large ensemble of image configurations, each one corresponding to a solution of the tomographic problem. In the case of IRET, the term "target" is used to represent a  source of thermal radiation. In AMIAS and RISE, each of the numerous randomly generated solutions describing the characteristics of the target is assigned a weight quantifying its "goodness" of representing the projection data.   The image is reconstructed by statistically weighting the entire ensemble of the simulated solutions. The formulation of RISE and the mathematical concepts of AMIAS are presented elsewhere ~\cite{Alexandrou:2015,Markou:2018, Papanicolas:2012, papanicolas:2018}. In this work, we summarize the essential features of the method and the specific choices required to implement it in the IRET problem; they are  described in the next paragraphs. 

\begin{figure}[ht!]
\begin{center}
\includegraphics[width=0.99\columnwidth]{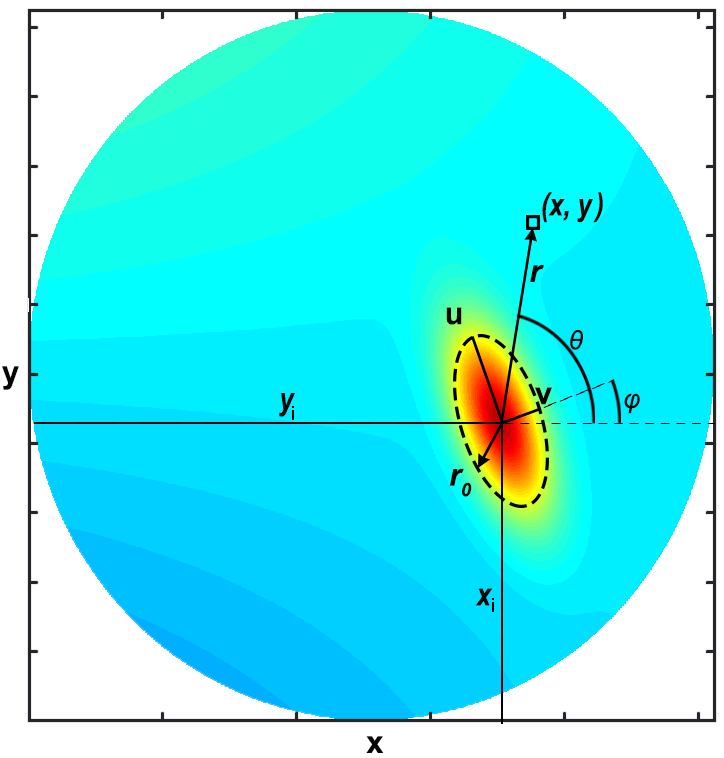}
\caption{The ellipsoidal model employed in the realization of RISE to represent the "elementary" shape of a thermal source. A sum of such elementary shapes superimposed on a smoothly varying background represents the distribution of temperature in the tomographic plane.  } 
\label{fig:model}
\end{center}
\end{figure}

\subsubsection{Modeling the Imaged Object}

 A model parametrized by a set of parameters is chosen to represent the physical characteristics of the target.  Prior knowledge can be incorporated at this step to endow the scheme with Bayesian capabilities.
 The model employed in RISE  to describe the temperature distribution is that of a summation of elementary sources (shown in Figure \ref{fig:model}) having an  ellipsoidal shape and embedded in a slowly varying background:
 
\begin{equation}
T(x,y) = \sum^{M_z}_{i=0}\sum^i_{j=-i}C^j_iZ^j_{i}(x,y)+\sum_{i=1}^{N_s}A_i(exp(\frac{r_i-r_{0i}}{s_ir_{0i}})+1)^{-1}
\label{Eq:Model}
\end{equation}
 where:
 \begin{itemize}
 
  \item {$Z^j_{i}(x,y)$ is a set of Zernike polynomials~\cite{len:2009,born:1959} representing the background tempereture distribution  and $C^j_i$ are their associated amplitudes,}
  \item{$M_z$ defines the total number of such polynomial required to approximate the background distribution,}
  \item {$N_s$ is the total number of elementary ellipsoidal targets,}
  \item {$A_i$ is the amplitude of temperature at the center of the $i^{th}$ target,} \item{$r_{i}$ is the euclidean distance of a point $(x,y)$ lying on the tomographic plane from the center $(x_i,y_i)$ of the $i^{th}$ target,}
  \item{ $s_i$  is a coefficient defining the "sharpness" of the temperature distribution in the surrounding by the source medium, and,}
 \item{$r_0(\theta)$  is a geometrical factor given as a function of the semi-major $u$ and semi-minor $v$ axes of an  ellipse respectively:
\begin{equation}
r_0(\theta) = \frac{u\cdot v}{\sqrt{(v \cdot cos(\phi-\theta))^2+ (u \cdot  sin(\phi - \theta))^2}}
\label{Eq:Ellipse}
\end{equation}
where the angle $\phi$ is the above parametric equation defines the orientation of the target in the tomographic plane.}
\end{itemize}

\subsubsection{Composing the Ensemble of Solutions} 

A Monte-Carlo procedure is used to sample the parameters of interest. Each set of sampled parameter values constituting a solution to the tomographic problem is used in  Equation \ref{Eq:Model} to construct a  tomographic image configuration. 
A set of projections is simulated from the tomographic image by employing the forward projection model presented in Equation \ref{Eq:Atten_Proj}. For each solution, the model estimated projections are compared to the measured projections through the $\chi^2$ criterion:
\begin{equation}
\chi^2 = \sum_{i}\frac{({{R}}_i-Y_i)^2}{\epsilon_i^2}
\label{Eq:x2}
\end{equation}
where $\epsilon_i$ is the associated error of the $i^{th}$ projection measurement.
The $\chi^2$ value is assigned to the solution to quantify its goodness of representing the data. The procedure is repeated to construct a large ensemble of possible solutions (typically 10$^5$ to 10$^6$), which as prescribed by the RISE method, is statistically weighted to determine the parameters of interest.

 One of the distinct features of RISE and the underlying AMIAS, which sets it apart from the standard reconstruction methods, is that it can incorporate any forward model simulating the propagation of the radiative energy from the source to the detector. 
This forward model can be a  function or a functional describing a very complex propagation process such as the entire cascade of non-uniform absorption and re-scattering.

\subsubsection{Determining the parameters of interest}

The RISE reconstruction process allows through different techniques \cite{Alexandrou:2015,Markou:2018, papanicolas:2018} the determination of the total number of terms ($N_s$)  which are needed in the model (Equation \ref{Eq:Model}) to represent the thermal sources.  The number $N^{opt}_s$ best describing the data is selected as the one minimizing the Bayesian Information Criterion (BIC)\cite{neath:2012}  given by the formula:
\begin{equation}
 \textrm{BIC} = \chi^2_{min}(N_s) + (7\cdot N_s+1) \cdot log{(M)}
 \label{eq:bic}
 \end{equation}
 where the $\chi^2_{min}(N_s)$ is the minimum $\chi^2$ value  in the ensemble of solutions constructed for the model of $N_s$ terms and $M$ is the number of ray projections used for the reconstruction of the tomographic image.

 Having defined the number $N^{opt}_s$,  the solutions  in the corresponding ensemble are  weighted with the probability value $exp(-\frac{1}{2}\chi^2)$ to derive the Probability Distribution Functions (PDFs) of the  parameters  $(A_i,x_{i},y_{i},u_{i},v_{i},\phi_{i},s_i)$. 
 Mean values and associated uncertainties are derived from the PDFs and used in Equation \ref{Eq:Model} to reconstruct the tomographic image of the temperature distribution.

\section{Simulation Studies} 
\label{Simulation Studies}

The validation of IRET and in particular the success of RISE in IRET has been accomplished with simulated data from numerical phantoms and real data from an experimental study with a hardware phantom. 

In three case studies employing five software phantoms, the ability of ART, MLEM, and RISE to reconstruct the image of the "true" temperature distribution has been examined using simulated sets of noisy infrared projections.

\subsection{Image Quality Metrics}
The objective assessment of a reconstruction method requires the calculation of different metrics quantifying the quality of the produced images. For the simulation studies presented, the metrics employed for this task are the Correlation Coefficient (CC), the Normalized Mean Square Error (NMSE), the Peak Signal-to-Noise Ratio (PSNR), the Contrast-to-Noise Ratio (CNR) and the Structural SIMilarity (SSIM) index. 

CC, which  provides a spatial similarity measure between two images \cite{Hill:2001}, is given by:
\begin{equation}
CC = \frac{\sum_{i=1}^{N^2}(F^a_i-\bar{F^a})(F^b_i-\bar{F^b})}{\sqrt{\sum_{i=1}^{N^2}(F^a_i-\bar{F}^a)^2\sum_{i=1}^{N^2}(F^b_i-\bar{F}^b)^2}}
\label{eq:corr}
\end{equation}
where $N^2$ is the total number of image pixels, $F^a_i$ and $F^b_i$ are the pixel values of the reconstructed and "true" image respectively, and $\bar{F^a}$, $\bar{F^b}$ are their corresponding average pixel values.  The CC as a metric expressing the  spatial similarity between two images it was used to evaluate the capability of ART, MLEM and RISE  to adequately resolve the geometrical characteristics of the imaged "targets".   

The NMSE and  PSNR metrics were used  to calculate the overall reconstruction error and to provide, respectively, a global measure of image contrast. The two metrics are defined as:
\begin{equation}
NMSE = \frac{\sum_{i=1}^{N^2}(F^a_i-F^b_i)^2}{\sum_{i=1}^{N^2}{F^b_i}^2}
\label{eq:nmse} 
\end{equation}
\begin{equation}
PSNR = 10log_{10}\bigg(\frac{N^2\cdot max(F^b)^2}{\sum_{i=1}^{N^2}(F^a_i-F^b_i)^2}\bigg)
\label{eq:psnr} 
\end{equation}

CNR is meant to objectively evaluate the detectability of a "target"  in a noisy background. It provides a local measure of contrast; it was calculated on a Region of Interest (ROI) extracted from the reconstructed image through the formula: 
\begin{equation}
CNR = \frac{\bar{T}-\bar{B}}{\sigma_B}
\label{eq:cnr} 
\end{equation}
where $\bar{T}$ is the average reconstructed temperature value in the target ROI, $\bar{B}$ and $\sigma_B$ is the average and the standard deviation, respectively, of image elements corresponding to the background ROI.

The SSIM index  is calculated on the reconstructed images with respect to the "true" phantom images to quantify their visual similarity.  SSIM is defined as a multiplicative combination of three indexes, namely the luminance index ($l$), the contrast index ($c$) and the similarity index ($s$) \cite{wang:2004}:
\begin{equation}
SSIM = l\cdot c \cdot s
\label{eq:ssim} 
\end{equation}
where the indexes $l$, $c$ and $s$ are defined in terms of the average pixel values  $\bar{F^a}$ and $\bar{F^b}$, the standard deviations $\sigma_a$ and $\sigma_b$, and the covariance $\sigma_{ab}$ of the reconstructed ($a$) and "true" ($b$) images:
\begin{equation}
l = \frac{2 \bar{F}^a \bar{F}^b +C_1}{{\bar{F^a}}^2+{\bar{F^b}}^2+C_1}
\label{eq:l} 
\end{equation}

\begin{equation}
c = \frac{2 \sigma_{a} \sigma_{b} +C_2}{\sigma^2_{a}+\sigma^2_{b} +C_2}
\label{eq:c} 
\end{equation}

\begin{equation}
s = \frac{\sigma_{ab}+C_3}{\sigma_a\sigma_b+C_3}
\label{eq:s} 
\end{equation}
The coefficients $C_1$, $C_2$ and $C_3$ are constants  introduced to prevent numerical instabilities when the denominators in the above equations are close to zero \cite{wang:2004}. Unlike the CC index evaluating the spatial characteristics of an image, the SSIM index accounts for the additional comparisons of pixel intensities to quantify the similarity in luminance and contrast between the reconstructed and "true" image.  

\begin{figure*}[ht!]
\begin{center}
\includegraphics[width=1.7\columnwidth]{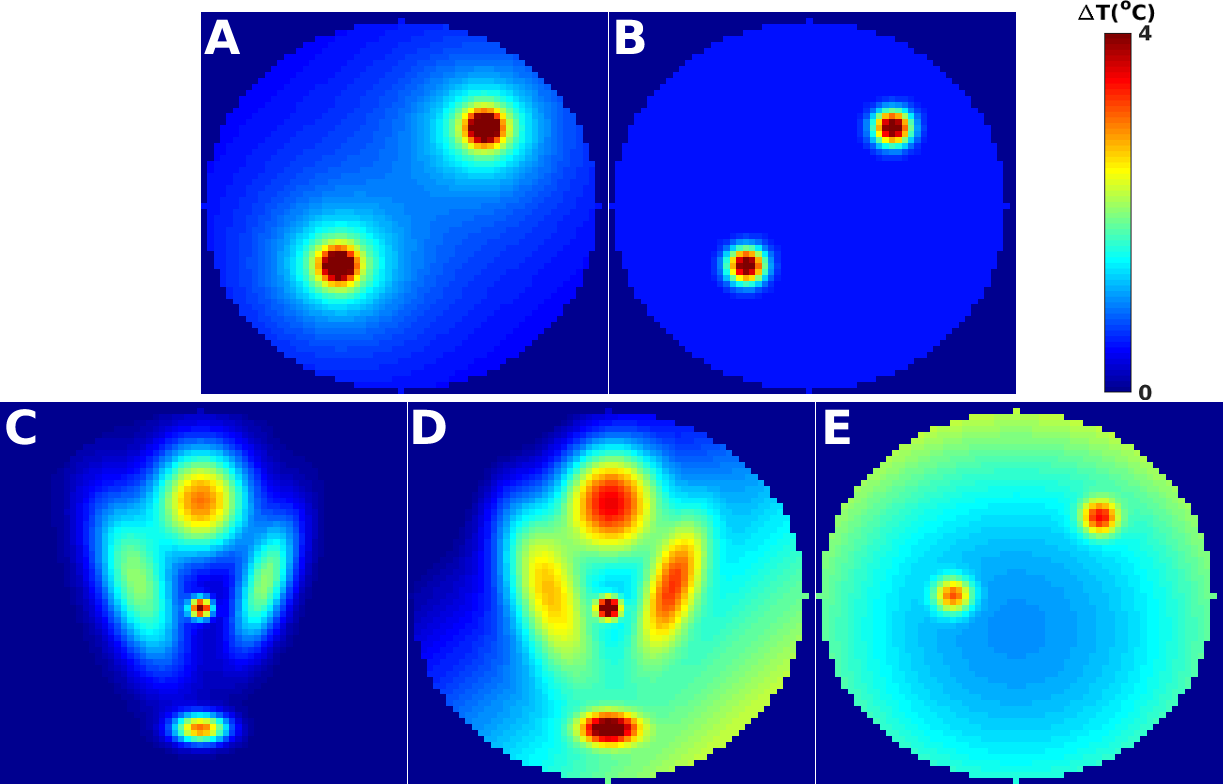}
\caption{The five numerical phantoms used in the simulation studies to evaluate different aspects of the RISE reconstruction methodology.} 
\label{fig:Phantom}
\end{center}
\end{figure*}
\begin{table*}[ht!]
\renewcommand{\arraystretch}{1.5} 
\caption{Descriptors of the Shepp-Logan variant - "Phantom C" - shown in Figure \ref{fig:Phantom}.}
\label{Tab:PhantomC}
\centering
\begin{tabular*}{1.9\columnwidth}{@{}c|@{\extracolsep{\fill}}cccc@{}}
\hline\hline
  
"Object"   & Temperature $T_0$  & Position $(x,y)$  & Size $(u,v)$  & Orientation $\theta$ \\ 
&  ($^{\circ}C$)  &   (pixel units)   &   (pixel units)  &  (degrees) \\ \hline
a  & 2  & (42.2, 34.0)  &  (7.9, 2.8) & 72 \\   
b  & 2  & (22.0, 34.0)  & (10.5, 4.1) & 108 \\
c  & 3  & (32.1, 48.0)  & (6.4, 5.4) & 0 \\
d  & 4  & (32.1, 30.0)  & (1.2, 1.2) & - \\
e  & 3  & (32.1, 9.8)   & (3.0, 1.7) & 0 \\
\hline\hline
\end{tabular*}
\end{table*}

\subsection{Numerical Phantoms and Projection Data Generation}
Five numerical  phantoms, shown in Figure \ref{fig:Phantom}, were used to generate the sets of infrared projections. The phantoms simulating  the presence of thermal sources ("hotspots") were defined on rectangular grids of $64\times 64$ pixels.

In the first phantom  ("A"), the  two depicted sources were generated using the radial temperature distribution:
 \begin{equation}
T(r) = T_0 \cdot \bigg(\theta(R_{0}-r) + \theta(r-R_{0})\cdot \frac{R_{0}}{r}\bigg)
\label{eq:PhantomA}
\end{equation}
where $T_0$ is a constant coefficient representing the temperature within the area of each hotspot, $R_0$ is the radius of the hotspot, and:
\[
 \theta(x) =
  \begin{cases}
   1 , & \quad \text{for } x \geq 0\\
   0 , & \quad \text{for
    }x < 0\\
  \end{cases}\] 
Phantom "B" has the same geometry as phantom "A"; however, the radial temperature distribution in the medium surrounding each  source exhibits a Gaussian profile, given by the equation: 
\begin{equation}
T(r) = T_0 \cdot exp\big(-\frac{1}{2}\frac{r^2}{R^2_{0}}\big) 
\label{eq:PhantomB}
\end{equation}
where, in this radial profile, the constant coefficient $T_0$ represents the temperature at the center of the hotspot and $R_0$ is  a constant defining the diffusion of temperature in the surrounding medium.
For both phantoms ("A" and "B"),  the coefficient $R_0$  was set to $2$ pixel units. 

Phantom "C" is a variant of the Shepp-Logan mathematical phantom; it consists five "hotspots" of different size and orientation having a Gaussian radial profile (see Equation \ref{eq:PhantomB}).  The parameters values defining the five hotspots are given in Table \ref{Tab:PhantomC}. 

Phantom "D" consists the same five hotspots of phantom "C" embedded in a disk having a non-uniform and non-symmetric temperature distribution. The temperature distribution of the disk introducing physical effects of background into the projection data was defined as a second order polynomial of the $(x,y)$ coordinates. The maximum temperature of the background was set to 2.2$^{\circ}$C.
Both "C" and "D" phantoms present challenging cases having five hotspots of different size, temperature, and orientation. Given that the projection data were generated by simulating attenuation conditions, the reconstruction of all hotspots and especially of the one in the middle, lying between the two larger ones, is particularly difficult.

Phantom "E" was also generated to present a non-uniform - non-symmetric temperature distribution. It presents two small hotspots, each one having a Gaussian radial profile (Equation \ref{eq:PhantomB}) and a radius of 2 pixel units. The temperature $T_0$ of the two hotspots was set to 2$^{\circ}$C, whereas the maximum temperature of the background was equal to 2.2$^{\circ}$C.

Sets of 24 infrared projections were generated from each phantom  using the forward projection model of Equation \ref{Eq:Atten_Proj}. 
For each simulation case, the set of 24 projections was generated in the full $360^\circ$ angular range with a constant angular step of $15^\circ$.  The detected infrared radiation $R_i$ resulting from the true temperature distribution $T_j$  was  calculated using the linearization of the problem  as provided in Equation \ref{Eq:Atten_Proj} for the spectral range  $[7.5\mu m,13\mu m]$   assuming that $\epsilon(\lambda)=1$. The uniform attenuation coefficient of the medium $k_u$ was set to the value of $0.1/w_p$ where $w_p$ is the size of a pixel. Each  projection was generated as a vector of $91$ rays $(NR=91)$ and  further randomized with Gaussian distributed noise $(n \sim \mathcal{N}(0,0.1^2))$. 

\begin{figure*}[ht!]
\begin{center}
\includegraphics[width=0.97\textwidth]{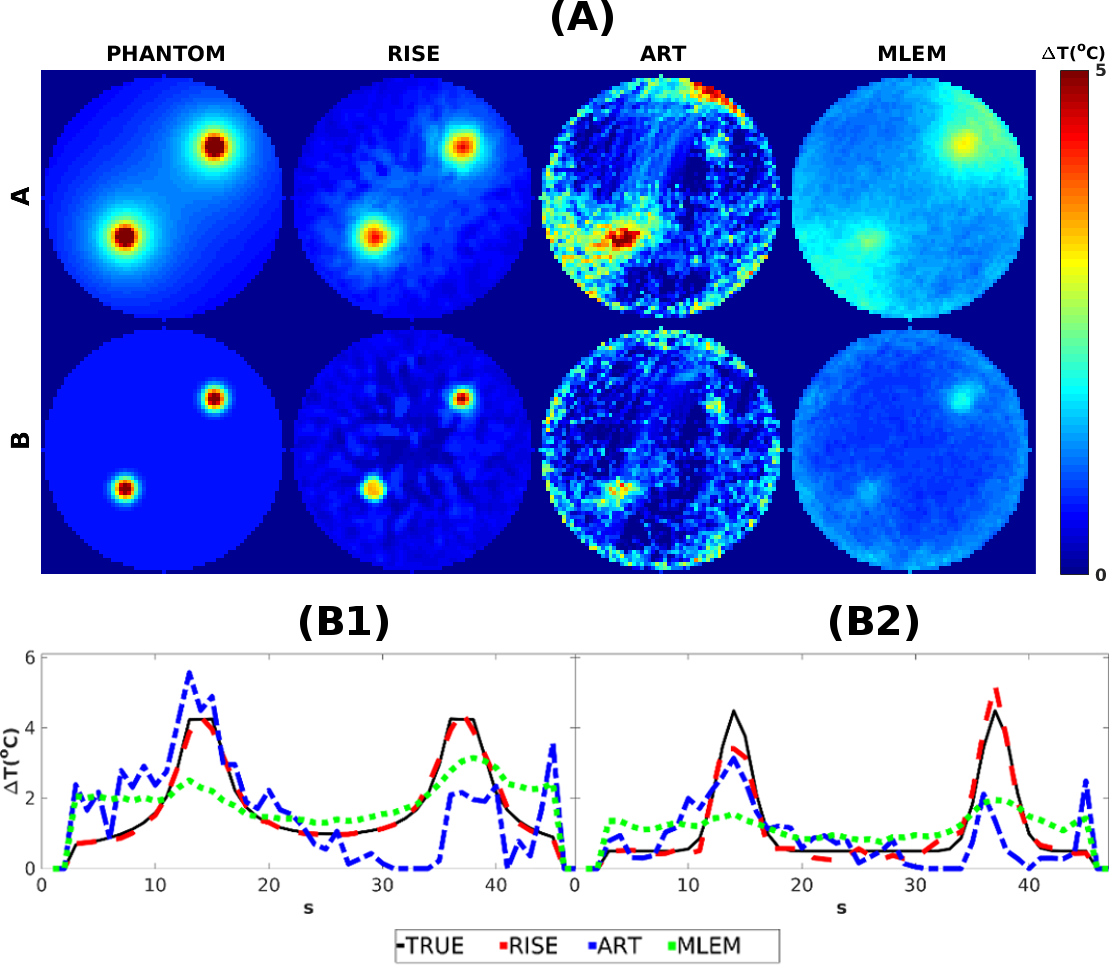}
\caption{(A) Reconstructed images of the phantoms "A" and "B" obtained in the first simulation study ("Model Capacity") with RISE, ART, and MLEM.  The temperature profiles drew from the reconstructed images  across the diagonal line connecting the centers of the two  hotspots are shown for  phantom "A" (B1) and  phantom "B" (B2).} 
\label{fig:Recs1}
\end{center}
\end{figure*}

The simulated sets of  projections were used in three distinct studies examining ART, MLEM and especially RISE in IRET:
\begin{enumerate}
\renewcommand{\labelenumi}{\Alph{enumi}}
\item{\textit{Model Capacity}: The model formulated in RISE (Equation \ref{Eq:Model}) is examined in its capability to provide sufficient reconstruction results when the "true" temperature distribution has a general and not necessarily the same radial temperature profile. Both of the two phantoms ("A", "B") defined  by setting the temperature coefficient $T_0$ to $4^oC$ were used in this study.}
\item{\textit{Resolving of Temperature Differences}: The efficacy of ART, MLEM, and RISE to reconstruct images revealing small differences in the temperature distribution is examined by the use of the phantom "A". In three simulation cases, the temperature coefficient $T_0$ was varied from $1^oC$ to $3^oC$ with a step size of $1^oC$  to produce different realizations of the temperature distribution. }

\item{\textit{Non-uniform Background Distribution}: This study was conducted to examine the capability of the three reconstruction methods to identify hotspots in the strong background. The effects of the non-uniform - non-symmetric background distribution on the reconstruction quality were examined using the numerical phantoms "C" and "D". The phantom "E", also presenting a non-uniform - non-symmetric background distribution was used to assess the detectability of small-sized targets in the produced reconstructions.}
\end{enumerate}

\subsection{Reconstruction Results}

\subsubsection{Model Capacity}

Figure \ref{fig:Recs1} shows the reconstructed images of the two simulated phantoms as they were produced in the first simulation study with RISE, ART and MLEM.  ART and MLEM reconstructions were obtained by performing two and three cycles of iterations respectively.  
\begin{table}[ht!]
\renewcommand{\arraystretch}{1.3} 
\caption{CC, NMSE, SSIM, PSNR and CNR scores of the reconstructed images presented in Figure \ref{fig:Recs1}.}
\label{tab:Recs_Model}
\centering
\begin{tabular*}{\columnwidth}{@{}l|@{\extracolsep{\fill}}ccc|ccc@{}}
\hline\hline
      & & Phantom A & & &Phantom B  & \\ 
	  & RISE   & ART   & MLEM   & RISE   & ART   & MLEM \\ \hline
CC    & 0.99   & 0.55  & 0.79   & 0.95   & 0.52  & 0.70 \\
NMSE  & 0.01   & 0.81  & 0.49   & 0.05   & 1.19  & 0.75 \\
SSIM  & 0.77   & 0.30  & 0.44   & 0.55   & 0.22  & 0.39 \\
PSNR  & 33.25  & 16.53 &12.78   & 32.33  & 16.41 & 11.24\\
CNR   & 6.52   & 2.24  & 3.17   & 13.55  & 1.97  & 3.11 \\
\hline\hline
\end{tabular*}
\end{table}

\begin{figure*}[ht!]
\centerline{
\includegraphics[width=0.97\textwidth]{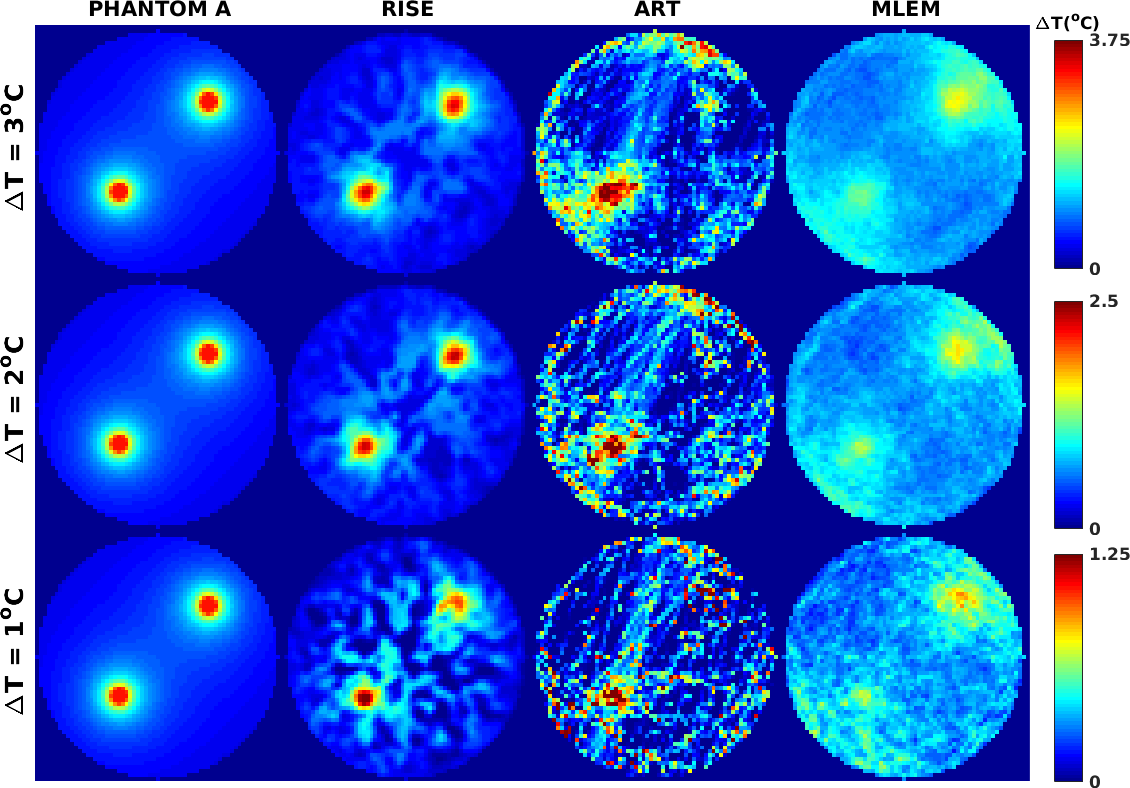}}
\caption{Image reconstructions obtained from the simulated projection data using RISE, ART and MLEM. The simulated phantoms used to generate the noisy sinograms are shown in the first column of the figure.}
\label{fig:Results0}
\end{figure*}

 From a visual inspection of the reconstructed images, it can be seen that RISE images exhibit higher contrast and less noise as compared to ART and MLEM reconstructions. Line-profiles extracted from the reconstructed images across the diagonal line connecting the centers of the two hotspots are depicted in the same figure.  As seen, the Fermi-like model employed in RISE yielded images that well-resolve the two simulated temperature distributions.

The image quality metrics (CC, NMSE, PSNR, CNR, and SSIM) comparing the reconstructed images are presented in Table \ref{tab:Recs_Model}. The results confirm and quantify the visual observations. For both  phantoms, while all the methods lead to acceptable image quality metrics, RISE led to superior CC and SSIM values (the highest among the three methods) indicating its ability to resolve the image of the temperature distribution adequately.   Moreover, the RISE images show an improved contrast and hotspots detectability as indicated by the calculated PSNR and CNR scores.

\subsubsection{Resolving of Temperature Differences}

  RISE, ART and MLEM were used to reconstruct the images of the phantom "A" for  three simulation cases varying the temperature coefficient $T_0$ in Equation \ref{eq:PhantomA} ($T_0=1^oC$, $T_0=2^oC$, $T_0=3^oC$). RISE reconstructions were produced using the Fermi-like model, the appropriateness of which is validated in the previous simulation study. For the case of ART and MLEM, the reconstructions were obtained by performing two and three grand iterations respectively.

The reconstructed images from the three methods are shown in Figure \ref{fig:Results0}. In all three simulation cases, streak artifacts reducing the detectability of the target are presented in ART reconstructions. RISE and MLEM images exhibit less amount of noise and provide reconstructions better revealing the simulated distribution of temperature. By visually examining these images it can be seen that the boundaries of the two hotspots are well shown in RISE and MLEM images, whereas, it is hard to detect them in the image produced with ART.

\begin{table}[h!]
\renewcommand{\arraystretch}{1.3} 
\caption{CC, NMSE, SSIM, PSNR and CNR scores of the RISE, ART, and MLEM images (shown in Figure \ref{fig:Results0}) reconstructed in the second simulation study ("Resolving of Temperature Differences"). }
\label{tab:Recs_Eval}
\centering
\begin{tabular*}{\columnwidth}{@{}ll@{\extracolsep{\fill}}cccc@{}}
\hline\hline
$T_0$ $(^oC)$        &     & RISE & ART & MLEM  \\ \hline
3.0    & CC & 0.96 & 0.55 & 0.80 \\ 
       & NMSE & 0.03 & 0.89 & 0.47  \\
       & SSIM & 0.65 & 0.22 & 0.45 \\
       & PSNR & 28.60 & 16.93 & 12.90 \\ 
       & CNR & 6.32 & 2.33 & 3.30 \\ \hline
2.0    & CC & 0.95 & 0.48 & 0.80  \\
       & NMSE & 0.05 & 1.18 & 0.48 \\
       & SSIM & 0.62 & 0.21 & 0.43 \\
       & PSNR & 26.78 & 17.95 & 13.24 \\ 
       & CNR & 6.28 & 2.07 & 3.41 \\ \hline
1.0    & CC & 0.86 & 0.46 & 0.77  \\
       & NMSE & 0.16 & 1.35 & 0.53 \\
       & SSIM & 0.52 & 0.22 & 0.42 \\
       & PSNR & 23.80 & 18.98 & 14.01 \\ 
       & CNR  & 5.04 & 1.71 & 2.50       \\      
\hline\hline
\end{tabular*}
\end{table}

The CC, NMSE, PSNR, CNR and SSIM scores evaluating the quality of the reconstructed images by the three methods are shown in Table \ref{tab:Recs_Eval}.
The calculated scores indicate superior image quality for the images reconstructed with RISE. As expected, the reconstruction quality of the three methods is degraded as the temperature difference $T_0$ in the simulated phantom is decreased. For the case exhibiting the smallest difference in temperature ($T_0 =1^oC$), the case with the most interest in this simulation study,  the highest CC and SSIM values obtained for  RISE  show an improvement in image similarity as compared to the corresponding CC and SSIM scores of MLEM and ART images respectively. 
For the same simulation case, the RISE reconstruction is shown to be superior concerning the hotspots detectability (CNR). RISE led to a CNR which is  $192\%$ and $100\%$ higher than those calculated for the images reconstructed with ART and MLEM  respectively. Likewise, the NMSE value obtained for RISE is more than two times less than those obtained for ART and MLEM. 

\subsubsection{Non-uniform Background Distribution}

\begin{figure*}[ht!]
\centerline{
\includegraphics[width=0.97\textwidth]{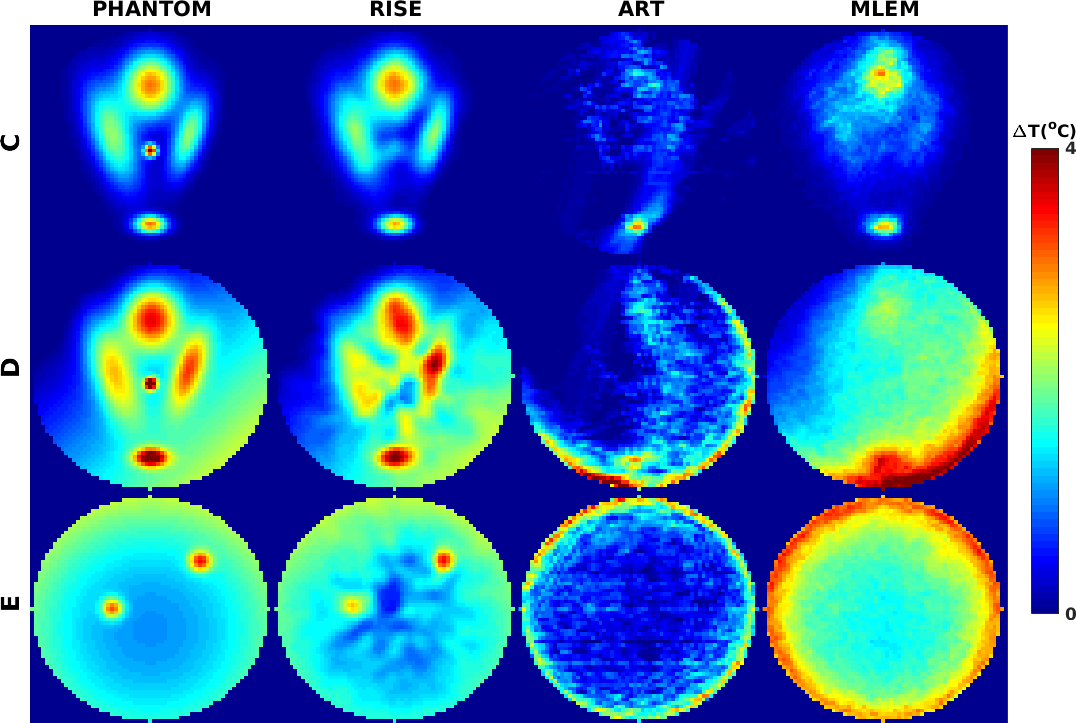}}
\caption{Images reconstructed from the simulated projection data using RISE, ART, and MLEM. The phantoms used for the simulation of the noisy sinograms are shown in the first column of the figure.}
\label{fig:ResultsBack}
\end{figure*}

Figure \ref{fig:ResultsBack} compares the reconstructed images of the phantoms "C", "D", and "E" obtained with RISE, ART and MLEM in the third simulation study. The calculated metrics (CC, NMSE, PSNR, CNR, SSIM) comparing the reconstructed images of the three methods are shown in Table \ref{tab:Recs_Var}.

In the absence of background (phantom "C"), as indicated by the CC and SSIM scores, the RISE image shows the highest spatial and structural similarity with the true image. In this "ideal" case of zero background, RISE presents improved image contrast and hotspots detectability as quantified via the PSNR and CNR respectively.
 From Figure \ref{fig:ResultsBack}, it can be seen that the reconstruction quality of the three methods is affected by the introduction of the non-uniform background in phantom "D". As the image quality metrics also indicate it in Table \ref{tab:Recs_Var}, both the structural similarity (SSIM) and the hotspots detectability (CNR) are reduced for all three methods. However, as compared to   ART and MLEM, the scores obtained for RISE show a superior performance which leads to an acceptable hotspots detectability (CNR) and a sufficiently high structural similarity (SSIM). Visually, the four of the five hotspots of "Phantom D" can be identified in the RISE image, whereas it is hard to be separated from the background in the images produced with ART and MLEM.
 
Differences in image characteristics between the three methods are also apparent in the reconstructions of phantom "E".  The combination of the non-uniform background with the small size of the hotspots simulated in this case makes the reconstruction problem difficult. As seen in Figure \ref{fig:ResultsBack}, the two hotspots cannot be observed in ART and MLEM reconstructions. The RISE image clearly shows the boundaries of the two hotspots which can be easily separated from the background. From Table \ref{tab:Recs_Var}, it can be seen that RISE yielded the highest spatial (CC) and structural (SSIM) similarity scores while it also led to the highest contrast (PSNR) and hotspots detectability (CNR).

Overall, the results from the three simulation studies validate the reconstruction efficacy of  RISE and indicate good performance in cases of non-uniform - non-symmetric background distributions.   The model which is employed in the method (Equation \ref{Eq:Model}) to represent the imaged targets proves to be adequate to describe hotspots exhibiting different radial intensity profiles. This feature of the method proves to be beneficial in reducing the amount of noise and improving the hotspots detectability of the reconstructed image.

\begin{table}[hb!]
\renewcommand{\arraystretch}{1.3} 
\caption{CC, NMSE, SSIM, PSNR, and CNR scores of the reconstructed images presented in Figure \ref{fig:ResultsBack}.}
\label{tab:Recs_Var}
\centering
\begin{tabular*}{\columnwidth}{@{}ll@{\extracolsep{\fill}}cccc@{}}
\hline\hline
Phantom       &   & RISE & ART & MLEM   \\ \hline
C & CC & 0.99 & 0.67 & 0.92  \\
& NMSE & 0.02 & 0.50 & 0.13 \\
& SSIM & 0.94 & 0.42 & 0.60 \\
& PSNR & 29.68 & 15.31 & 21.24\\ 
& CNR & 5.39 & 1.66 & 3.28 \\ \hline
D & CC & 0.98 & 0.56 & 0.83  \\
& NMSE & 0.01 & 0.46 & 0.18 \\
& SSIM & 0.82 & 0.36 & 0.49 \\
& PSNR & 27.51 & 12.64 & 16.76\\ 
& CNR & 2.00 & 0.18 & 0.08 \\ \hline
E & CC & 0.99 & 0.73 & 0.96  \\
& NMSE & 0.01 & 0.27 & 0.20 \\
& SSIM & 0.77 & 0.44 & 0.66 \\
& PSNR & 29.46 & 13.98 & 15.31\\ 
& CNR & 3.68 & 0.33 & 0.43 \\
\hline\hline
\end{tabular*}
\end{table}

\begin{figure*}[ht!]
\centering{
\includegraphics[width=1.75\columnwidth]{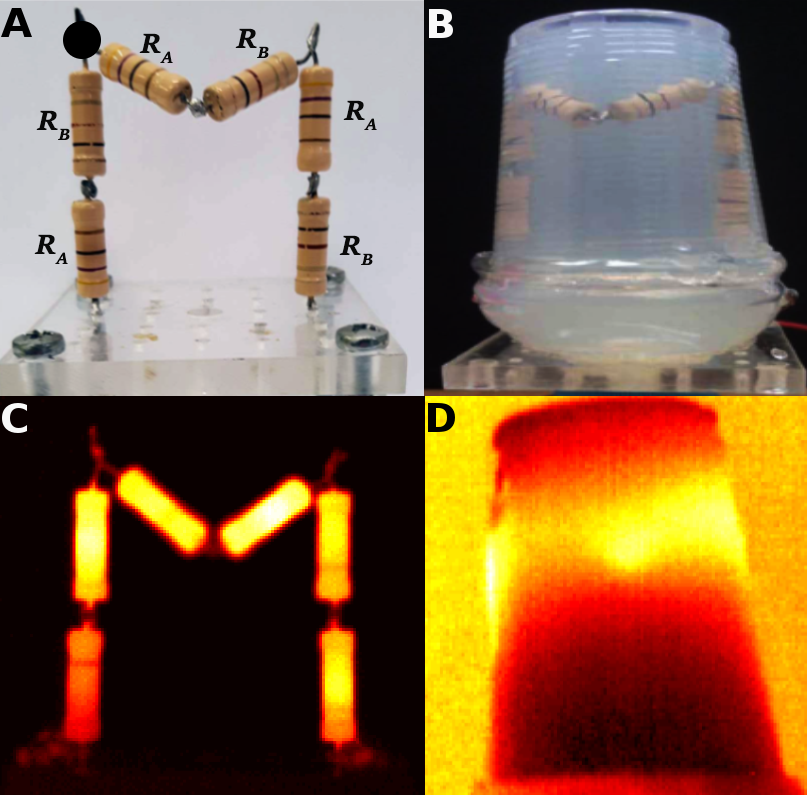}}
\caption{(A) The thermal phantom constructed as a configuration of six resistors forming the deformed capital letter 'M'. The temperature of the phantom was monitored by attaching a thermocouple at the point indicated by the black circle. (B) The phantom was submerged in absorbing gel and imaged with an infrared camera from a distance of 15 cm. Planar images (thermographs) of the phantom captured in the zero absorption (C) and absorption (D) experimental studies.}
\label{Fig:Phantom}
\end{figure*}

\section{Experiments with a Thermal Phantom}
\label{Thermal Phantom}

The evaluation of RISE in IRET  was further explored using sets of thermal images captured from a thermal phantom. In two experimental studies reported in \cite{Rapsomanikis:2012}, sets of 24 infrared images were acquired from the phantom using an infrared camera. In the present study, RISE and the two conventional methods ART and MLEM  are examined in their capacity to reconstruct the tomographic images of the phantom temperature distribution. The derived tomographs by the three methods were assembled in stacks to visualize the 3D  distribution of temperature in the volume of interest.

\subsection{The Thermal Phantom}

The thermal phantom was constructed out of three pairs of $R_A=100 \Omega$ and $R_B=47 \Omega$ resistors, alternately configured to form the capital letter 'M', with the two middle resistors oriented out of the plane (Figure \ref{Fig:Phantom}). The resistors were connected in series and supplied with direct current. A thermocouple monitoring the temperature of the phantom was adjusted at the upper-left corner of the configuration as it is shown in Figure \ref{Fig:Phantom}.

Two experimental studies~\cite{Rapsomanikis:2012} were carried out with the implemented hardware phantom:
\begin{enumerate}
\renewcommand{\labelenumi}{\Alph{enumi}}
\item \textit{Zero Absorption Case}. The thermal radiation was captured as it was emitted directly from the resistors with no intervening medium. The temperature of the phantom measured by a thermocouple was comparable to the core temperature of humans ($38^oC$), and the environmental temperature measured at a distance of 5 cm from the phantom was  $25^oC$. The set of infrared projections obtained in this study was used to provide reconstructed images for visual comparisons.
\item \textit{Absorption Case}. The same configuration of six resistors was placed in a conical vessel filled with agarose gel of about $1$ gr/ml concentration. The inner diameter of the vessel was $40$ mm (small diameter) at its top surface and $65$ mm at its bottom surface (big diameter). The environmental temperature was  $25^oC$, and so was the temperature of the upper-left resistor,  kept at $38^oC$.  In this study, we examine the ability of RISE  to reconstruct tomographic images in absorption conditions degrading the quality of the planar thermal images.
\end{enumerate}

\subsection{Data Acquisition}

In both experiments, the phantom was placed on a  rotating table at a distance of $15$ cm from the front of the camera. A set of $24$ infrared images of the phantom, captured using a thermal camera (Thermovision 570, AGEMA Infrared Systems),  were obtained in the full range of $360^{\circ}$  with an angular step of $15^{\circ}$.  The thermal camera has a $24^o \times 18^o$ Field of View (FOV) and is characterized by $0.1^oC$ thermal sensitivity; it operates in the spectral range of 7.5 to 13 $\mu m$. The 24 planar images (thermographs) were further sliced to provide a set of  15 sinograms, each one corresponding to a tomographic level at a specific vertical offset. The extracted sinograms were used as inputs in RISE, ART, and MLEM to reconstruct the sets of $64\times64$ tomographic images of the phantom. 
\begin{figure*}[ht!]
\begin{center}
\includegraphics[width=0.97\textwidth]{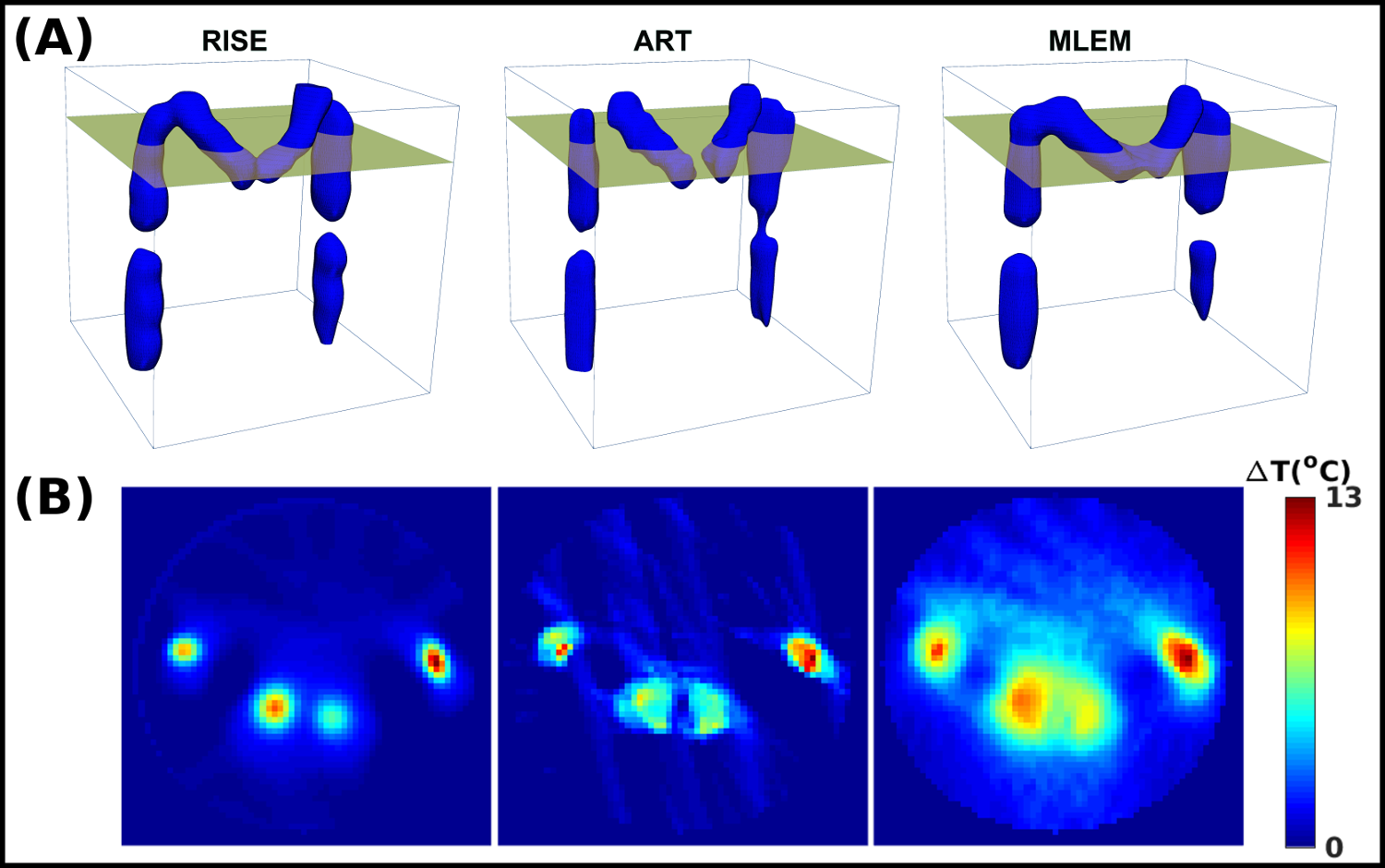}\\
\caption{(A) The 3D reconstructions of the phantom obtained in the first experimental study (nil absorption conditions) with RISE, ART, and MLEM. (B) Reconstructed tomographic images corresponding to the sectional plane indicated in the 3D reconstructions with yellow color.}
\label{Fig:Results1}
\end{center}
\end{figure*}

\subsection{Reconstruction Results}

\begin{figure*}[ht!]
\begin{center}
\includegraphics[width=0.97\textwidth]{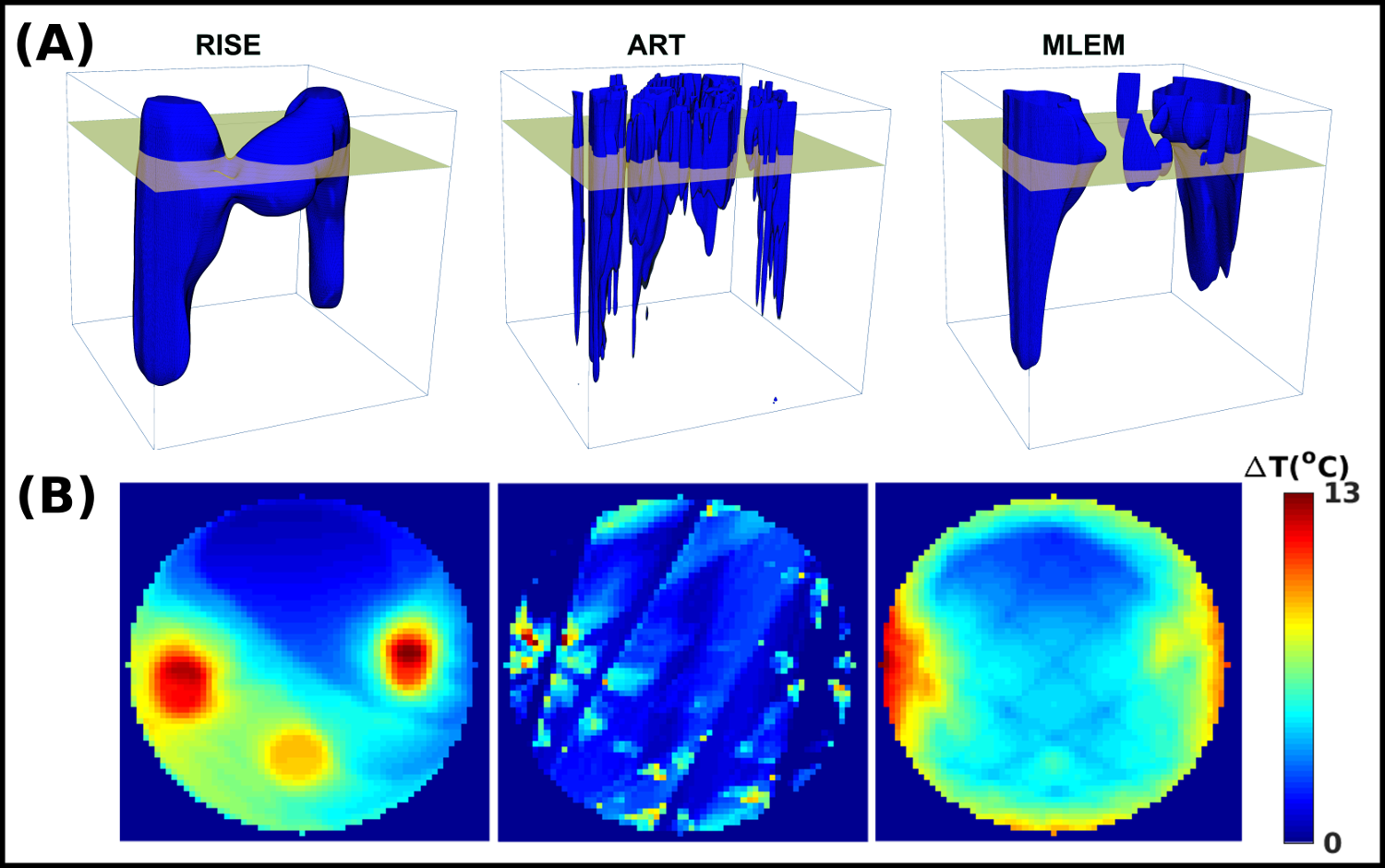}
\caption{(A) The 3D reconstructions of the thermal phantom produced with RISE, ART and MLEM in the second experimental study introducing absorption conditions. (B) Reconstructed images  corresponding to the tomographic planes indicated by the yellow color as obtained with the three methods. }
\label{Fig:Results2}
\end{center}
\end{figure*}

The reconstructed images obtained in the two experimental studies, with and without absorption conditions, are visually compared in Figures \ref{Fig:Results1} and \ref{Fig:Results2}. The capability of each method to reproduce the structure of the imaged 'M' shaped object and to visualize the difference in temperature distribution from the alternatively positioned resistors were the two primary criteria for the evaluation of the images.

All the three methods yielded reliable reconstructions in the absorption free case (Figure \ref{Fig:Results1}) validating the IRET methodology as described in Section \ref{IRET Methodology}.   From a visual inspection of the tomographic images reconstructed by the three methods (shown at the bottom panel of Figure \ref{Fig:Results1}), it can be seen that the alternation in temperature between the $100\Omega$ and $47\Omega$ resistors is visualized in the RISE image. This alternation in temperature cannot be observed in the images reconstructed with ART or MLEM.

The   RISE, ART and MLEM 3D volumetric images of the phantom (shown at the top panel of Figure \ref{Fig:Results1}) obtained by assembling the 2D tomographic images in stacks indicate that all three methods successfully resolved the geometry of the phantom.
In the case of RISE, the structure of the phantom was visualized by surfacing the volumetric data at the determined model radius parameter $r_0$ (see Equation \ref{Eq:Ellipse}). In the case of ART and MLEM, isothermal surfaces were extracted at half of the maximum temperature to resolve the 3D shape of the thermal phantom.

    The tomographic images of the phantom produced in the second experimental study ("Absorption Case") with the three methods are shown in  Figure \ref{Fig:Results2}. For all three reconstruction methods, images were obtained by using  a uniform attenuation coefficient $\kappa_u= 0.23 mm^{-1}$ (Equation \ref{Eq:Atten_Proj}). From a visual inspection of the images produced in this case study, it can be observed that the results obtained with RISE are superior. ART images exhibit noisy artifacts and provide limited information about the structure of the phantom. Compared to MLEM, the RISE images exhibit higher contrast and provide better approximations of the thermal sources. In the images produced by the three methods, the difference in temperature from the alternatively positioned resistors is not observed.
    
\begin{figure}[ht!]
\begin{center}
\includegraphics[width=\columnwidth]{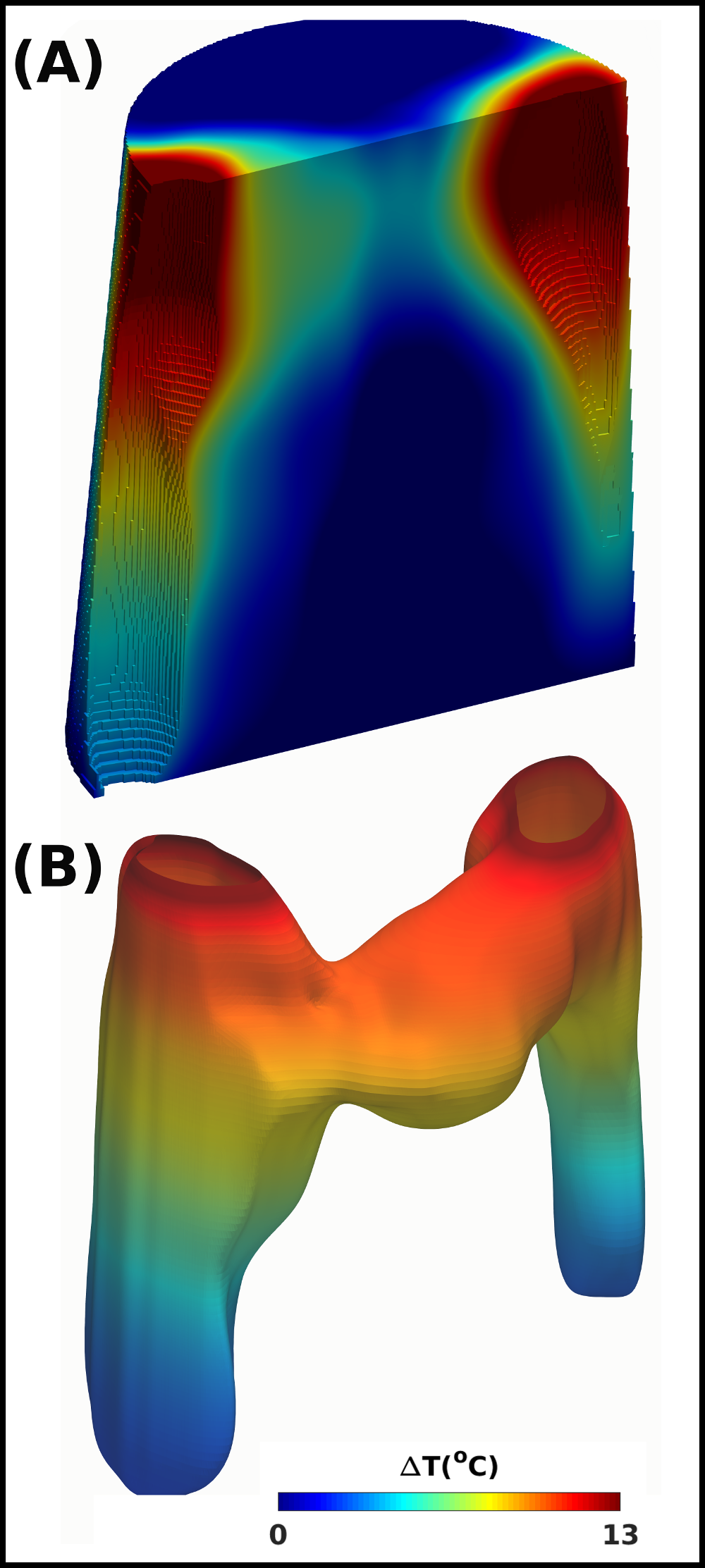}
\end{center}
\caption{ The volumetric representation of the temperature distribution of the phantom was produced by assembling the reconstructed tomographic images in a stack. The  3D images were rendered in ParaView \cite{Henderson:2004} using surfacing and contouring visualization techniques. (A) A vertical section of the entire volume of the phantom; (B) The distribution of temperature on the surface of the thermal heaters.}
\label{Fig:amias3d}
\end{figure}    

    The  3D images of the phantom composed from the sets of the reconstructed images by following the same procedure used in the previous case ("Zero Absorption Case") are shown at the top panel of Figure \ref{Fig:Results2}.   RISE yields more reliable imaging of the 'M' shaped phantom and to a lesser degree  MLEM, while ART fails. The amplitude of temperature at the center of the phantom, indicating the presence of the middle pair of resistors, is more visible in the RISE result.

  An additional 3D image of the phantom visualizing the  RISE reconstruction is shown in Figure \ref{Fig:amias3d}. This volumetric image depicts the distribution of temperature on the adjacent surface lying at a distance $r_0$ from the center of sources. The 3D reconstruction reveals a positive gradient in temperature which is observed across the vertical axis ($z$-axis). This gradient is caused by the difference in the produced thermal energy between the upper and bottom part of the phantom, comprising four and two heaters respectively. It can also be observed that compared to the RISE image obtained in the  "Zero Absorption Case," the image obtained in this study with the same method presents an over-sized representation of the thermal sources.  This difference can be understood as resulting from the heated medium (agarose gel) surrounding the resistors. A more precise image of the heaters, if such a result is desired, would require the implementation of a more sophisticated model that takes into account heat transport effects. Such a forward model can be implemented in the general framework of RISE to accommodate the diffusion of infrared radiation within the absorbing material.

The tomographic images of the hardware phantom produced by RISE  compared to those of MLEM and ART, demonstrate the suitability of RISE for Infrared Tomography. These images should also be compared with the planar thermal images (thermographs) of the phantom (shown in panel D of Figure \ref{Fig:Phantom}), which do not allow any conclusions to be drawn for the thermal distribution within the gel.

\section{Conclusions}
\label{Conclusions}
In the work presented in this paper, the Infrared Emission Tomography (IRET) technique is revisited, and it is shown that the use of the methodology employed in ionizing radiation tomography (e.g., PET or SPECT modalities) is both justified and implementable.  The IRET methodology implemented with the widely used MLEM and ART techniques yields satisfactory results, especially in cases where attenuation and scattering effects are minimal. The RISE technique yields superior tomographic images even in cases where medium modifications are present.

   Images from all three reconstruction methods were evaluated using well-established metrics of the field.  All yielded acceptable tomographic results according to these metrics although RISE images exhibit a significantly higher structural similarity (SSIM) to the simulated distributions and improved hotspot detectability (CNR). The three methods were also used to reconstruct tomographic images from experimental data obtained from a thermal phantom imaged in nil and high absorption conditions.  In the case of absorption, which represents a semi-realistic case simulating attenuation conditions in the human body, RISE yielded an image that reliably resolves the geometry of the phantom and visualizes the gradient of the temperature distribution within the image medium. 
 
The robust methodology presented in this work specifies how further refinements can be accomplished with the inclusion of detailed modeling of non-uniform attenuation and scattering effects. Following the reported successful implementation of the RISE method in IRET further expanded experimentation with phantoms and possibly small animals is warranted to ascertain the potential of this modality in medical imaging.

\begin{acknowledgements}
This  work was supported by the Graduate School of The Cyprus Institute and the Cy-Tera Project "NEA IPODOMI/STRATI/0308/31", which is co-funded by the European Regional Development Fund and the Republic of Cyprus through the Research Promotion Foundation.
\end{acknowledgements}

\section*{Conflicts of interest}
We have no conflict of interest to declare.

\bibliography{bibliog}

\begin{thebibliography}{47}%
\makeatletter
\providecommand \@ifxundefined [1]{%
 \@ifx{#1\undefined}
}%
\providecommand \@ifnum [1]{%
 \ifnum #1\expandafter \@firstoftwo
 \else \expandafter \@secondoftwo
 \fi
}%
\providecommand \@ifx [1]{%
 \ifx #1\expandafter \@firstoftwo
 \else \expandafter \@secondoftwo
 \fi
}%
\providecommand \natexlab [1]{#1}%
\providecommand \enquote  [1]{``#1''}%
\providecommand \bibnamefont  [1]{#1}%
\providecommand \bibfnamefont [1]{#1}%
\providecommand \citenamefont [1]{#1}%
\providecommand \href@noop [0]{\@secondoftwo}%
\providecommand \href [0]{\begingroup \@sanitize@url \@href}%
\providecommand \@href[1]{\@@startlink{#1}\@@href}%
\providecommand \@@href[1]{\endgroup#1\@@endlink}%
\providecommand \@sanitize@url [0]{\catcode `\\12\catcode `\$12\catcode
  `\&12\catcode `\#12\catcode `\^12\catcode `\_12\catcode `\%12\relax}%
\providecommand \@@startlink[1]{}%
\providecommand \@@endlink[0]{}%
\providecommand \url  [0]{\begingroup\@sanitize@url \@url }%
\providecommand \@url [1]{\endgroup\@href {#1}{\urlprefix }}%
\providecommand \urlprefix  [0]{URL }%
\providecommand \Eprint [0]{\href }%
\providecommand \doibase [0]{http://dx.doi.org/}%
\providecommand \selectlanguage [0]{\@gobble}%
\providecommand \bibinfo  [0]{\@secondoftwo}%
\providecommand \bibfield  [0]{\@secondoftwo}%
\providecommand \translation [1]{[#1]}%
\providecommand \BibitemOpen [0]{}%
\providecommand \bibitemStop [0]{}%
\providecommand \bibitemNoStop [0]{.\EOS\space}%
\providecommand \EOS [0]{\spacefactor3000\relax}%
\providecommand \BibitemShut  [1]{\csname bibitem#1\endcsname}%
\let\auto@bib@innerbib\@empty
\bibitem [{\citenamefont {Ring}\ and\ \citenamefont {Ammer}(2012)}]{ring:2012}%
  \BibitemOpen
  \bibfield  {author} {\bibinfo {author} {\bibfnamefont {E.}~\bibnamefont
  {Ring}}\ and\ \bibinfo {author} {\bibfnamefont {K.}~\bibnamefont {Ammer}},\
  }\href@noop {} {\bibfield  {journal} {\bibinfo  {journal} {Physiol. Meas.}\
  }\textbf {\bibinfo {volume} {33}},\ \bibinfo {pages} {R33} (\bibinfo {year}
  {2012})}\BibitemShut {NoStop}%
\bibitem [{\citenamefont {Baddour}(2006)}]{Bad:2006}%
  \BibitemOpen
  \bibfield  {author} {\bibinfo {author} {\bibfnamefont {N.}~\bibnamefont
  {Baddour}},\ }\href@noop {} {\bibfield  {journal} {\bibinfo  {journal} {J.
  Phys. A Math. Theor.}\ }\textbf {\bibinfo {volume} {39}},\ \bibinfo {pages}
  {14379} (\bibinfo {year} {2006})}\BibitemShut {NoStop}%
\bibitem [{\citenamefont {Lahiri}\ \emph {et~al.}(2012)\citenamefont {Lahiri},
  \citenamefont {Bagavathiappan}, \citenamefont {Jayakumar},\ and\
  \citenamefont {Philip}}]{lahiri:2012}%
  \BibitemOpen
  \bibfield  {author} {\bibinfo {author} {\bibfnamefont {B.}~\bibnamefont
  {Lahiri}}, \bibinfo {author} {\bibfnamefont {S.}~\bibnamefont
  {Bagavathiappan}}, \bibinfo {author} {\bibfnamefont {T.}~\bibnamefont
  {Jayakumar}}, \ and\ \bibinfo {author} {\bibfnamefont {J.}~\bibnamefont
  {Philip}},\ }\href@noop {} {\bibfield  {journal} {\bibinfo  {journal}
  {Infrared Phys. Technol.}\ }\textbf {\bibinfo {volume} {55}},\ \bibinfo
  {pages} {221} (\bibinfo {year} {2012})}\BibitemShut {NoStop}%
\bibitem [{\citenamefont {Keyserlingk}\ \emph {et~al.}(2000)\citenamefont
  {Keyserlingk} \emph {et~al.}}]{keyser:2000}%
  \BibitemOpen
  \bibfield  {author} {\bibinfo {author} {\bibfnamefont {J.}~\bibnamefont
  {Keyserlingk}} \emph {et~al.},\ }\href@noop {} {\bibfield  {journal}
  {\bibinfo  {journal} {{IEEE} Eng. Med. Biol. Mag.}\ }\textbf {\bibinfo
  {volume} {19}},\ \bibinfo {pages} {30} (\bibinfo {year} {2000})}\BibitemShut
  {NoStop}%
\bibitem [{\citenamefont {Ring}\ and\ \citenamefont
  {Collins}(1970)}]{ring:1970}%
  \BibitemOpen
  \bibfield  {author} {\bibinfo {author} {\bibfnamefont {E.}~\bibnamefont
  {Ring}}\ and\ \bibinfo {author} {\bibfnamefont {A.}~\bibnamefont {Collins}},\
  }\href@noop {} {\bibfield  {journal} {\bibinfo  {journal} {Rheumatology}\
  }\textbf {\bibinfo {volume} {10}},\ \bibinfo {pages} {337} (\bibinfo {year}
  {1970})}\BibitemShut {NoStop}%
\bibitem [{\citenamefont {Glehr}\ \emph {et~al.}(2011)\citenamefont {Glehr}
  \emph {et~al.}}]{glehr:2011}%
  \BibitemOpen
  \bibfield  {author} {\bibinfo {author} {\bibfnamefont {M.}~\bibnamefont
  {Glehr}} \emph {et~al.},\ }\href@noop {} {\bibfield  {journal} {\bibinfo
  {journal} {Int. J. Thermodynamics}\ }\textbf {\bibinfo {volume} {14}},\
  \bibinfo {pages} {71} (\bibinfo {year} {2011})}\BibitemShut {NoStop}%
\bibitem [{\citenamefont {Romano}\ \emph {et~al.}(2011)\citenamefont {Romano}
  \emph {et~al.}}]{romano:2011}%
  \BibitemOpen
  \bibfield  {author} {\bibinfo {author} {\bibfnamefont {C.~L.}\ \bibnamefont
  {Romano}} \emph {et~al.},\ }\href@noop {} {\bibfield  {journal} {\bibinfo
  {journal} {J. Orthop. Traumatol.}\ }\textbf {\bibinfo {volume} {12}},\
  \bibinfo {pages} {81} (\bibinfo {year} {2011})}\BibitemShut {NoStop}%
\bibitem [{\citenamefont {Varju}\ \emph {et~al.}(2004)\citenamefont {Varju}
  \emph {et~al.}}]{varju:2004}%
  \BibitemOpen
  \bibfield  {author} {\bibinfo {author} {\bibfnamefont {G.}~\bibnamefont
  {Varju}} \emph {et~al.},\ }\href@noop {} {\bibfield  {journal} {\bibinfo
  {journal} {Rheumatology}\ }\textbf {\bibinfo {volume} {43}},\ \bibinfo
  {pages} {915} (\bibinfo {year} {2004})}\BibitemShut {NoStop}%
\bibitem [{\citenamefont {Mayr}(1995)}]{mayr:1995}%
  \BibitemOpen
  \bibfield  {author} {\bibinfo {author} {\bibfnamefont {H.}~\bibnamefont
  {Mayr}},\ }\href@noop {} {\bibfield  {journal} {\bibinfo  {journal} {The
  Thermal Image in Medicine and Biology}\ ,\ \bibinfo {pages} {182}} (\bibinfo
  {year} {1995})}\BibitemShut {NoStop}%
\bibitem [{\citenamefont {Vardasca}(2012)}]{vardasca:2012}%
  \BibitemOpen
  \bibfield  {author} {\bibinfo {author} {\bibfnamefont {R.~{\^A}.~R.}\
  \bibnamefont {Vardasca}},\ }\emph {\bibinfo {title} {The effect of work
  related mechanical stress on the peripheral temperature of the hand}},\
  \href@noop {} {Ph.D. thesis},\ \bibinfo  {school} {University of Glamorgan}
  (\bibinfo {year} {2012})\BibitemShut {NoStop}%
\bibitem [{\citenamefont {Herman}\ and\ \citenamefont
  {Cetingul}(2011)}]{herman:2011}%
  \BibitemOpen
  \bibfield  {author} {\bibinfo {author} {\bibfnamefont {C.}~\bibnamefont
  {Herman}}\ and\ \bibinfo {author} {\bibfnamefont {M.~P.}\ \bibnamefont
  {Cetingul}},\ }\href@noop {} {\bibfield  {journal} {\bibinfo  {journal} {J.
  Vis. Exp.}\ }\textbf {\bibinfo {volume} {51}},\ \bibinfo {pages} {e2679}
  (\bibinfo {year} {2011})}\BibitemShut {NoStop}%
\bibitem [{\citenamefont {Cruz}\ \emph {et~al.}(2009)\citenamefont {Cruz} \emph
  {et~al.}}]{santa:2009}%
  \BibitemOpen
  \bibfield  {author} {\bibinfo {author} {\bibfnamefont {S.}~\bibnamefont
  {Cruz}} \emph {et~al.},\ }\href@noop {} {\bibfield  {journal} {\bibinfo
  {journal} {Appl. Radiat. Isot.}\ }\textbf {\bibinfo {volume} {67}},\ \bibinfo
  {pages} {S54} (\bibinfo {year} {2009})}\BibitemShut {NoStop}%
\bibitem [{\citenamefont {Maillard}\ and\ \citenamefont
  {Hessler}(1969)}]{maillard:1969}%
  \BibitemOpen
  \bibfield  {author} {\bibinfo {author} {\bibfnamefont {G.}~\bibnamefont
  {Maillard}}\ and\ \bibinfo {author} {\bibfnamefont {C.}~\bibnamefont
  {Hessler}},\ }\href@noop {} {\bibfield  {journal} {\bibinfo  {journal}
  {Dermatologica}\ }\textbf {\bibinfo {volume} {139}},\ \bibinfo {pages} {353}
  (\bibinfo {year} {1969})}\BibitemShut {NoStop}%
\bibitem [{\citenamefont {Kirubha}\ \emph {et~al.}(2015)\citenamefont
  {Kirubha}, \citenamefont {Anburajan}, \citenamefont {Venkataraman},\ and\
  \citenamefont {Menaka}}]{kirubha:2015}%
  \BibitemOpen
  \bibfield  {author} {\bibinfo {author} {\bibfnamefont {A.~S.}\ \bibnamefont
  {Kirubha}}, \bibinfo {author} {\bibfnamefont {M.}~\bibnamefont {Anburajan}},
  \bibinfo {author} {\bibfnamefont {B.}~\bibnamefont {Venkataraman}}, \ and\
  \bibinfo {author} {\bibfnamefont {M.}~\bibnamefont {Menaka}},\ }\href@noop {}
  {\bibfield  {journal} {\bibinfo  {journal} {Infrared Phys. Technol.}\
  }\textbf {\bibinfo {volume} {73}},\ \bibinfo {pages} {115} (\bibinfo {year}
  {2015})}\BibitemShut {NoStop}%
\bibitem [{\citenamefont {Kennedy}\ \emph {et~al.}(2009)\citenamefont
  {Kennedy}, \citenamefont {Lee},\ and\ \citenamefont {Seely}}]{kennedy:2009}%
  \BibitemOpen
  \bibfield  {author} {\bibinfo {author} {\bibfnamefont {D.~A.}\ \bibnamefont
  {Kennedy}}, \bibinfo {author} {\bibfnamefont {T.}~\bibnamefont {Lee}}, \ and\
  \bibinfo {author} {\bibfnamefont {D.}~\bibnamefont {Seely}},\ }\href@noop {}
  {\bibfield  {journal} {\bibinfo  {journal} {Integr. Cancer Ther.}\ }\textbf
  {\bibinfo {volume} {8}},\ \bibinfo {pages} {9} (\bibinfo {year}
  {2009})}\BibitemShut {NoStop}%
\bibitem [{\citenamefont {Ng}\ and\ \citenamefont {Kee}(2008)}]{ng:2008}%
  \BibitemOpen
  \bibfield  {author} {\bibinfo {author} {\bibfnamefont {E.}~\bibnamefont
  {Ng}}\ and\ \bibinfo {author} {\bibfnamefont {E.}~\bibnamefont {Kee}},\
  }\href@noop {} {\bibfield  {journal} {\bibinfo  {journal} {J. Med. Eng.
  Technol.}\ }\textbf {\bibinfo {volume} {32}},\ \bibinfo {pages} {103}
  (\bibinfo {year} {2008})}\BibitemShut {NoStop}%
\bibitem [{\citenamefont {Lapayowker}\ \emph {et~al.}(1973)\citenamefont
  {Lapayowker}, \citenamefont {Salen}, \citenamefont {Ziskin},\ and\
  \citenamefont {Rosemond}}]{lapa:1973}%
  \BibitemOpen
  \bibfield  {author} {\bibinfo {author} {\bibfnamefont {M.~S.}\ \bibnamefont
  {Lapayowker}}, \bibinfo {author} {\bibfnamefont {S.}~\bibnamefont {Salen}},
  \bibinfo {author} {\bibfnamefont {M.}~\bibnamefont {Ziskin}}, \ and\ \bibinfo
  {author} {\bibfnamefont {G.~P.}\ \bibnamefont {Rosemond}},\ }\href@noop {}
  {\bibfield  {journal} {\bibinfo  {journal} {Cancer}\ }\textbf {\bibinfo
  {volume} {31}},\ \bibinfo {pages} {777} (\bibinfo {year} {1973})}\BibitemShut
  {NoStop}%
\bibitem [{\citenamefont {Cetingul}\ and\ \citenamefont
  {Herman}(2010)}]{Pirtini:2010}%
  \BibitemOpen
  \bibfield  {author} {\bibinfo {author} {\bibfnamefont {M.~P.}\ \bibnamefont
  {Cetingul}}\ and\ \bibinfo {author} {\bibfnamefont {C.}~\bibnamefont
  {Herman}},\ }\href@noop {} {\bibfield  {journal} {\bibinfo  {journal} {Phys.
  Med. Biol.}\ }\textbf {\bibinfo {volume} {55}},\ \bibinfo {pages} {5933}
  (\bibinfo {year} {2010})}\BibitemShut {NoStop}%
\bibitem [{\citenamefont {Song}\ \emph {et~al.}(1984)\citenamefont {Song} \emph
  {et~al.}}]{Song:1984}%
  \BibitemOpen
  \bibfield  {author} {\bibinfo {author} {\bibfnamefont {C.~W.}\ \bibnamefont
  {Song}} \emph {et~al.},\ }\href@noop {} {\bibfield  {journal} {\bibinfo
  {journal} {IEEE Trans. Biomed. Eng.}\ }\textbf {\bibinfo {volume} {BME-31}},\
  \bibinfo {pages} {9} (\bibinfo {year} {1984})}\BibitemShut {NoStop}%
\bibitem [{\citenamefont {Snyder}\ \emph {et~al.}(2000)\citenamefont {Snyder},
  \citenamefont {Qi}, \citenamefont {Elliott}, \citenamefont {Head},\ and\
  \citenamefont {Wang}}]{snyder:2000}%
  \BibitemOpen
  \bibfield  {author} {\bibinfo {author} {\bibfnamefont {W.~E.}\ \bibnamefont
  {Snyder}}, \bibinfo {author} {\bibfnamefont {H.}~\bibnamefont {Qi}}, \bibinfo
  {author} {\bibfnamefont {R.~L.}\ \bibnamefont {Elliott}}, \bibinfo {author}
  {\bibfnamefont {J.~F.}\ \bibnamefont {Head}}, \ and\ \bibinfo {author}
  {\bibfnamefont {C.~X.}\ \bibnamefont {Wang}},\ }\href@noop {} {\bibfield
  {journal} {\bibinfo  {journal} {{IEEE} Eng. Med. Biol. Mag}\ }\textbf
  {\bibinfo {volume} {19}},\ \bibinfo {pages} {63} (\bibinfo {year}
  {2000})}\BibitemShut {NoStop}%
\bibitem [{\citenamefont {Hossain}\ \emph {et~al.}(2013)\citenamefont
  {Hossain}, \citenamefont {Lu}, \citenamefont {Sun},\ and\ \citenamefont
  {Yan}}]{Hossain:2013}%
  \BibitemOpen
  \bibfield  {author} {\bibinfo {author} {\bibfnamefont {M.~M.}\ \bibnamefont
  {Hossain}}, \bibinfo {author} {\bibfnamefont {G.}~\bibnamefont {Lu}},
  \bibinfo {author} {\bibfnamefont {D.}~\bibnamefont {Sun}}, \ and\ \bibinfo
  {author} {\bibfnamefont {Y.}~\bibnamefont {Yan}},\ }\href@noop {} {\bibfield
  {journal} {\bibinfo  {journal} {Meas. Sci. Technol.}\ }\textbf {\bibinfo
  {volume} {24}},\ \bibinfo {pages} {074010} (\bibinfo {year}
  {2013})}\BibitemShut {NoStop}%
\bibitem [{\citenamefont {Goyal}\ \emph {et~al.}(2014)\citenamefont {Goyal},
  \citenamefont {Chaudhry},\ and\ \citenamefont {Subbarao}}]{Goyal:2014}%
  \BibitemOpen
  \bibfield  {author} {\bibinfo {author} {\bibfnamefont {A.}~\bibnamefont
  {Goyal}}, \bibinfo {author} {\bibfnamefont {S.}~\bibnamefont {Chaudhry}}, \
  and\ \bibinfo {author} {\bibfnamefont {P.}~\bibnamefont {Subbarao}},\
  }\href@noop {} {\bibfield  {journal} {\bibinfo  {journal} {Combust. Flame}\
  }\textbf {\bibinfo {volume} {161}},\ \bibinfo {pages} {173 } (\bibinfo {year}
  {2014})}\BibitemShut {NoStop}%
\bibitem [{\citenamefont {{Lienhard IV}}\ and\ \citenamefont {{Lienhard
  V}}(2008)}]{lienhard:2013}%
  \BibitemOpen
  \bibfield  {author} {\bibinfo {author} {\bibfnamefont {J.}~\bibnamefont
  {{Lienhard IV}}}\ and\ \bibinfo {author} {\bibfnamefont {J.}~\bibnamefont
  {{Lienhard V}}},\ }\href@noop {} {\emph {\bibinfo {title} {A Heat Transfer
  Textbook}}},\ \bibinfo {edition} {3rd}\ ed.\ (\bibinfo  {publisher}
  {Phlogiston Press},\ \bibinfo {address} {Cambridge, Massachusetts},\ \bibinfo
  {year} {2008})\ pp.\ \bibinfo {pages} {3--46}\BibitemShut {NoStop}%
\bibitem [{\citenamefont {Modest}(2013)}]{modest:2013}%
  \BibitemOpen
  \bibfield  {author} {\bibinfo {author} {\bibfnamefont {M.~F.}\ \bibnamefont
  {Modest}},\ }in\ \href@noop {} {\emph {\bibinfo {booktitle} {Radiative Heat
  Transfer}}},\ \bibinfo {editor} {edited by\ \bibinfo {editor} {\bibfnamefont
  {M.~F.}\ \bibnamefont {Modest}}}\ (\bibinfo  {publisher} {Academic Press},\
  \bibinfo {address} {Boston},\ \bibinfo {year} {2013})\ \bibinfo {edition}
  {3rd}\ ed.,\ pp.\ \bibinfo {pages} {1 -- 30}\BibitemShut {NoStop}%
\bibitem [{\citenamefont {Lalush}\ and\ \citenamefont
  {Wernick}(2004)}]{Wernick:2004}%
  \BibitemOpen
  \bibfield  {author} {\bibinfo {author} {\bibfnamefont {D.~S.}\ \bibnamefont
  {Lalush}}\ and\ \bibinfo {author} {\bibfnamefont {M.~N.}\ \bibnamefont
  {Wernick}},\ }in\ \href@noop {} {\emph {\bibinfo {booktitle} {Emission
  Tomography}}},\ \bibinfo {editor} {edited by\ \bibinfo {editor}
  {\bibfnamefont {M.~N.}\ \bibnamefont {Wernick}}\ and\ \bibinfo {editor}
  {\bibfnamefont {J.~N.}\ \bibnamefont {Aarsvold}}}\ (\bibinfo  {publisher}
  {Academic Press},\ \bibinfo {address} {San Diego},\ \bibinfo {year} {2004})\
  pp.\ \bibinfo {pages} {443 -- 472}\BibitemShut {NoStop}%
\bibitem [{\citenamefont {Natterer}\ and\ \citenamefont
  {Wübbeling}(2001)}]{Natterer:2001}%
  \BibitemOpen
  \bibfield  {author} {\bibinfo {author} {\bibfnamefont {F.}~\bibnamefont
  {Natterer}}\ and\ \bibinfo {author} {\bibfnamefont {F.}~\bibnamefont
  {Wübbeling}},\ }\href@noop {} {\emph {\bibinfo {title} {Mathematical Methods
  in Image Reconstruction}}}\ (\bibinfo  {publisher} {SIAM},\ \bibinfo
  {address} {Philadelphia},\ \bibinfo {year} {2001})\ pp.\ \bibinfo {pages}
  {44--45}\BibitemShut {NoStop}%
\bibitem [{\citenamefont {Ishimaru}(1978)}]{ishimaru:1978}%
  \BibitemOpen
  \bibfield  {author} {\bibinfo {author} {\bibfnamefont {A.}~\bibnamefont
  {Ishimaru}},\ }\href@noop {} {\emph {\bibinfo {title} {Wave propagation and
  scattering in random media}}}\ (\bibinfo  {publisher} {Academic Press},\
  \bibinfo {year} {1978})\BibitemShut {NoStop}%
\bibitem [{\citenamefont {Bohren}\ and\ \citenamefont
  {Huffman}(2008)}]{bohren:2008}%
  \BibitemOpen
  \bibfield  {author} {\bibinfo {author} {\bibfnamefont {C.~F.}\ \bibnamefont
  {Bohren}}\ and\ \bibinfo {author} {\bibfnamefont {D.~R.}\ \bibnamefont
  {Huffman}},\ }\href@noop {} {\emph {\bibinfo {title} {Absorption and
  scattering of light by small particles}}}\ (\bibinfo  {publisher} {John Wiley
  \& Sons},\ \bibinfo {address} {New York},\ \bibinfo {year}
  {2008})\BibitemShut {NoStop}%
\bibitem [{\citenamefont {Nissil{\"a}}\ \emph {et~al.}(2005)\citenamefont
  {Nissil{\"a}}, \citenamefont {Noponen}, \citenamefont {Heino}, \citenamefont
  {Kajava},\ and\ \citenamefont {Katila}}]{Nissila:2005}%
  \BibitemOpen
  \bibfield  {author} {\bibinfo {author} {\bibfnamefont {I.}~\bibnamefont
  {Nissil{\"a}}}, \bibinfo {author} {\bibfnamefont {T.}~\bibnamefont
  {Noponen}}, \bibinfo {author} {\bibfnamefont {J.}~\bibnamefont {Heino}},
  \bibinfo {author} {\bibfnamefont {T.}~\bibnamefont {Kajava}}, \ and\ \bibinfo
  {author} {\bibfnamefont {T.}~\bibnamefont {Katila}},\ }in\ \href@noop {}
  {\emph {\bibinfo {booktitle} {Advances in electromagnetic fields in living
  systems}}}\ (\bibinfo  {publisher} {Springer},\ \bibinfo {address} {Boston},\
  \bibinfo {year} {2005})\ pp.\ \bibinfo {pages} {77--129}\BibitemShut
  {NoStop}%
\bibitem [{\citenamefont {Gordon}\ \emph {et~al.}(1970)\citenamefont {Gordon}
  \emph {et~al.}}]{Gordon:1970}%
  \BibitemOpen
  \bibfield  {author} {\bibinfo {author} {\bibfnamefont {R.}~\bibnamefont
  {Gordon}} \emph {et~al.},\ }\href@noop {} {\bibfield  {journal} {\bibinfo
  {journal} {J. Theor. Biol.}\ }\textbf {\bibinfo {volume} {29}},\ \bibinfo
  {pages} {471 } (\bibinfo {year} {1970})}\BibitemShut {NoStop}%
\bibitem [{\citenamefont {Gilbert}(1972)}]{Gilbert:1972}%
  \BibitemOpen
  \bibfield  {author} {\bibinfo {author} {\bibfnamefont {P.}~\bibnamefont
  {Gilbert}},\ }\href@noop {} {\bibfield  {journal} {\bibinfo  {journal} {J.
  Theor. Biol.}\ }\textbf {\bibinfo {volume} {36}},\ \bibinfo {pages} {105 }
  (\bibinfo {year} {1972})}\BibitemShut {NoStop}%
\bibitem [{\citenamefont {Shepp}\ and\ \citenamefont
  {Vardi}(1982)}]{Shepp:1982}%
  \BibitemOpen
  \bibfield  {author} {\bibinfo {author} {\bibfnamefont {L.~A.}\ \bibnamefont
  {Shepp}}\ and\ \bibinfo {author} {\bibfnamefont {Y.}~\bibnamefont {Vardi}},\
  }\href@noop {} {\bibfield  {journal} {\bibinfo  {journal} {IEEE Trans. Med.
  Imaging}\ }\textbf {\bibinfo {volume} {1}},\ \bibinfo {pages} {113} (\bibinfo
  {year} {1982})}\BibitemShut {NoStop}%
\bibitem [{\citenamefont {Lange}\ \emph {et~al.}(1984)\citenamefont {Lange}
  \emph {et~al.}}]{Lange:1984}%
  \BibitemOpen
  \bibfield  {author} {\bibinfo {author} {\bibfnamefont {K.}~\bibnamefont
  {Lange}} \emph {et~al.},\ }\href@noop {} {\bibfield  {journal} {\bibinfo
  {journal} {J. Comput. Assist. Tomogr.}\ }\textbf {\bibinfo {volume} {8}},\
  \bibinfo {pages} {306} (\bibinfo {year} {1984})}\BibitemShut {NoStop}%
\bibitem [{\citenamefont {Wieczorek}(2010)}]{Wiez:2010}%
  \BibitemOpen
  \bibfield  {author} {\bibinfo {author} {\bibfnamefont {H.}~\bibnamefont
  {Wieczorek}},\ }\href@noop {} {\bibfield  {journal} {\bibinfo  {journal}
  {Phys. Med. Biol.}\ }\textbf {\bibinfo {volume} {55}},\ \bibinfo {pages}
  {3161} (\bibinfo {year} {2010})}\BibitemShut {NoStop}%
\bibitem [{\citenamefont {Angeli}\ and\ \citenamefont
  {Stiliaris}(2009)}]{Angeli:2009}%
  \BibitemOpen
  \bibfield  {author} {\bibinfo {author} {\bibfnamefont {S.}~\bibnamefont
  {Angeli}}\ and\ \bibinfo {author} {\bibfnamefont {E.}~\bibnamefont
  {Stiliaris}},\ }\href@noop {} {\bibfield  {journal} {\bibinfo  {journal}
  {IEEE NSS-MIC}\ ,\ \bibinfo {pages} {3382}} (\bibinfo {year}
  {2009})}\BibitemShut {NoStop}%
\bibitem [{\citenamefont {Hwang}\ and\ \citenamefont
  {Zeng}(2006)}]{Hwang:2006}%
  \BibitemOpen
  \bibfield  {author} {\bibinfo {author} {\bibfnamefont {D.}~\bibnamefont
  {Hwang}}\ and\ \bibinfo {author} {\bibfnamefont {G.~L.}\ \bibnamefont
  {Zeng}},\ }\href@noop {} {\bibfield  {journal} {\bibinfo  {journal} {Phys.
  Med. Biol.}\ }\textbf {\bibinfo {volume} {51}},\ \bibinfo {pages} {237}
  (\bibinfo {year} {2006})}\BibitemShut {NoStop}%
\bibitem [{\citenamefont {Alexandrou}\ \emph {et~al.}(2015)\citenamefont
  {Alexandrou}, \citenamefont {Leontiou}, \citenamefont {Papanicolas},\ and\
  \citenamefont {Stiliaris}}]{Alexandrou:2015}%
  \BibitemOpen
  \bibfield  {author} {\bibinfo {author} {\bibfnamefont {C.}~\bibnamefont
  {Alexandrou}}, \bibinfo {author} {\bibfnamefont {T.}~\bibnamefont
  {Leontiou}}, \bibinfo {author} {\bibfnamefont {C.~N.}\ \bibnamefont
  {Papanicolas}}, \ and\ \bibinfo {author} {\bibfnamefont {E.}~\bibnamefont
  {Stiliaris}},\ }\href@noop {} {\bibfield  {journal} {\bibinfo  {journal}
  {Phys. Rev. D}\ }\textbf {\bibinfo {volume} {91}},\ \bibinfo {pages} {014506}
  (\bibinfo {year} {2015})}\BibitemShut {NoStop}%
\bibitem [{\citenamefont {{L. Markou}}\ \emph {et~al.}(2018)\citenamefont {{L.
  Markou}}, \citenamefont {{E. Stiliaris}},\ and\ \citenamefont {{C. N.
  Papanicolas}}}]{Markou:2018}%
  \BibitemOpen
  \bibfield  {author} {\bibinfo {author} {\bibnamefont {{L. Markou}}}, \bibinfo
  {author} {\bibnamefont {{E. Stiliaris}}}, \ and\ \bibinfo {author}
  {\bibnamefont {{C. N. Papanicolas}}},\ }\href@noop {} {\bibfield  {journal}
  {\bibinfo  {journal} {Eur. Phys. J. A}\ }\textbf {\bibinfo {volume} {54}},\
  \bibinfo {pages} {115} (\bibinfo {year} {2018})}\BibitemShut {NoStop}%
\bibitem [{\citenamefont {Papanicolas}\ and\ \citenamefont
  {Stiliaris}(2012)}]{Papanicolas:2012}%
  \BibitemOpen
  \bibfield  {author} {\bibinfo {author} {\bibfnamefont {C.~N.}\ \bibnamefont
  {Papanicolas}}\ and\ \bibinfo {author} {\bibfnamefont {E.}~\bibnamefont
  {Stiliaris}},\ }\href@noop {} {\bibfield  {journal} {\bibinfo  {journal}
  {arXiv: 1205.6505}\ } (\bibinfo {year} {2012})}\BibitemShut {NoStop}%
\bibitem [{\citenamefont {{Papanicolas, C. N. and Koutsantonis, L. and
  Stiliaris, E.}}(2018)}]{papanicolas:2018}%
  \BibitemOpen
  \bibfield  {author} {\bibinfo {author} {\bibnamefont {{Papanicolas, C. N. and
  Koutsantonis, L. and Stiliaris, E.}}},\ }\href@noop {} {\bibfield  {journal}
  {\bibinfo  {journal} {arXiv: 1804.03915}\ } (\bibinfo {year}
  {2018})}\BibitemShut {NoStop}%
\bibitem [{\citenamefont {Lencrerot}\ \emph {et~al.}(2009)\citenamefont
  {Lencrerot}, \citenamefont {Litman}, \citenamefont {Tortel},\ and\
  \citenamefont {Geffrin}}]{len:2009}%
  \BibitemOpen
  \bibfield  {author} {\bibinfo {author} {\bibfnamefont {R.}~\bibnamefont
  {Lencrerot}}, \bibinfo {author} {\bibfnamefont {A.}~\bibnamefont {Litman}},
  \bibinfo {author} {\bibfnamefont {H.}~\bibnamefont {Tortel}}, \ and\ \bibinfo
  {author} {\bibfnamefont {J.-M.}\ \bibnamefont {Geffrin}},\ }\href@noop {}
  {\bibfield  {journal} {\bibinfo  {journal} {Inverse Probl.}\ }\textbf
  {\bibinfo {volume} {25}},\ \bibinfo {pages} {035012} (\bibinfo {year}
  {2009})}\BibitemShut {NoStop}%
\bibitem [{\citenamefont {Born}\ and\ \citenamefont {Wolf}(1959)}]{born:1959}%
  \BibitemOpen
  \bibfield  {author} {\bibinfo {author} {\bibfnamefont {M.}~\bibnamefont
  {Born}}\ and\ \bibinfo {author} {\bibfnamefont {E.}~\bibnamefont {Wolf}},\
  }\href@noop {} {\emph {\bibinfo {title} {Principles of optics}}}\ (\bibinfo
  {publisher} {Pergamon},\ \bibinfo {address} {New York},\ \bibinfo {year}
  {1959})\BibitemShut {NoStop}%
\bibitem [{\citenamefont {Neath}\ and\ \citenamefont
  {Cavanaugh}(2012)}]{neath:2012}%
  \BibitemOpen
  \bibfield  {author} {\bibinfo {author} {\bibfnamefont {A.~A.}\ \bibnamefont
  {Neath}}\ and\ \bibinfo {author} {\bibfnamefont {J.~E.}\ \bibnamefont
  {Cavanaugh}},\ }\href@noop {} {\bibfield  {journal} {\bibinfo  {journal}
  {Wiley Interdisciplinary Reviews: Computational Statistics}\ }\textbf
  {\bibinfo {volume} {4}},\ \bibinfo {pages} {199} (\bibinfo {year}
  {2012})}\BibitemShut {NoStop}%
\bibitem [{\citenamefont {Hill}\ \emph {et~al.}(2001)\citenamefont {Hill} \emph
  {et~al.}}]{Hill:2001}%
  \BibitemOpen
  \bibfield  {author} {\bibinfo {author} {\bibfnamefont {D.~L.~G.}\
  \bibnamefont {Hill}} \emph {et~al.},\ }\href@noop {} {\bibfield  {journal}
  {\bibinfo  {journal} {Phys. Med. Biol.}\ }\textbf {\bibinfo {volume} {46}},\
  \bibinfo {pages} {R1} (\bibinfo {year} {2001})}\BibitemShut {NoStop}%
\bibitem [{\citenamefont {Wang}\ \emph {et~al.}(2004)\citenamefont {Wang},
  \citenamefont {Bovik}, \citenamefont {Sheikh},\ and\ \citenamefont
  {Simoncelli}}]{wang:2004}%
  \BibitemOpen
  \bibfield  {author} {\bibinfo {author} {\bibfnamefont {Z.}~\bibnamefont
  {Wang}}, \bibinfo {author} {\bibfnamefont {A.~C.}\ \bibnamefont {Bovik}},
  \bibinfo {author} {\bibfnamefont {H.~R.}\ \bibnamefont {Sheikh}}, \ and\
  \bibinfo {author} {\bibfnamefont {E.~P.}\ \bibnamefont {Simoncelli}},\
  }\href@noop {} {\bibfield  {journal} {\bibinfo  {journal} {IEEE Tran. Image
  Process.}\ }\textbf {\bibinfo {volume} {13}},\ \bibinfo {pages} {600}
  (\bibinfo {year} {2004})}\BibitemShut {NoStop}%
\bibitem [{\citenamefont {Rapsomanikis}\ \emph {et~al.}(2012)\citenamefont
  {Rapsomanikis}, \citenamefont {Zioga}, \citenamefont {Kontos}, \citenamefont
  {Mikeli},\ and\ \citenamefont {Stiliaris}}]{Rapsomanikis:2012}%
  \BibitemOpen
  \bibfield  {author} {\bibinfo {author} {\bibfnamefont {A.~N.}\ \bibnamefont
  {Rapsomanikis}}, \bibinfo {author} {\bibfnamefont {M.}~\bibnamefont {Zioga}},
  \bibinfo {author} {\bibfnamefont {M.}~\bibnamefont {Kontos}}, \bibinfo
  {author} {\bibfnamefont {M.}~\bibnamefont {Mikeli}}, \ and\ \bibinfo {author}
  {\bibfnamefont {E.}~\bibnamefont {Stiliaris}},\ }\href@noop {} {\bibfield
  {journal} {\bibinfo  {journal} {IEEE NSS-MIC}\ ,\ \bibinfo {pages} {3632}}
  (\bibinfo {year} {2012})}\BibitemShut {NoStop}%
\bibitem [{\citenamefont {Henderson}\ \emph {et~al.}(2004)\citenamefont
  {Henderson}, \citenamefont {Ahrens}, \citenamefont {Law} \emph
  {et~al.}}]{Henderson:2004}%
  \BibitemOpen
  \bibfield  {author} {\bibinfo {author} {\bibfnamefont {A.}~\bibnamefont
  {Henderson}}, \bibinfo {author} {\bibfnamefont {J.}~\bibnamefont {Ahrens}},
  \bibinfo {author} {\bibfnamefont {C.}~\bibnamefont {Law}},  \emph {et~al.},\
  }\href@noop {} {\emph {\bibinfo {title} {The ParaView Guide}}},\ Vol.\
  \bibinfo {volume} {366}\ (\bibinfo  {publisher} {Kitware},\ \bibinfo
  {address} {Clifton Park, New York},\ \bibinfo {year} {2004})\BibitemShut
  {NoStop}%
\end{thebibliography}%

\end{document}